\documentclass[manuscript,trackchanges]{aastex631}
\usepackage{needspace}

\newcommand{\trot}{T_{\rm rot}}
\newcommand{\tkin}{T_{\rm kin}}
\newcommand{\fullpagefig}[1]{\includegraphics[width=\textwidth,height=0.82\textheight,keepaspectratio]{#1}}

\shorttitle{Warm Interstellar HD and H$_2$}
\shortauthors{Neufeld}

\begin{document}

\title{Excitation of warm HD and H$_2$ and implications for the interstellar deuterium abundance}

\received{June 22, 2026}
\accepted{June 30, 2026}

\correspondingauthor{David A. Neufeld}
\email{neufeld@jhu.edu}

\author[0000-0001-8341-1646]{David A. Neufeld}
\affiliation{William H. Miller III Department of Physics and Astronomy, Johns Hopkins University, Baltimore, MD 21218, USA}

\begin{abstract}
Rotational emission from HD provides a direct probe of deuterium in warm
molecular gas, but the conversion of HD line fluxes to total column densities
is sensitive to non-LTE excitation.  I present calculations of the non-LTE
rotational excitation of HD and H$_2$ and a method for fitting their rotational
diagrams simultaneously.  Because HD remains subthermally excited to higher
densities than H$_2$, the combined analysis constrains the thermal pressure as well
as the HD/H$_2$ abundance ratio.  I apply this method to JWST/MIRI observations
of 26 positions in 10 protostellar outflows from the JOYS survey, published
previously by Francis et al. (2025), including 17
positions with secure detections of at least three HD lines.  Two-temperature,
isobaric models generally provide good fits to the measured
H$_2$ and HD level populations.  For 11 positions with HD detections, the inferred
pressures span $p/k_B=8.4\times10^7$ -- $5.1\times10^9\ {\rm K\,cm^{-3}}$; for
the remaining six, the HD populations approach LTE and provide only lower limits
on the gas pressure.  Excluding two positions toward IRAS 4B with comparatively poor
fits, the median and mean inferred HD/H$_2$ abundance ratios are
$1.42\times10^{-5}$ and $1.47\times10^{-5}$, respectively, only 28\% and
29\% of the value expected if all primordial deuterium were present as
HD.  These results confirm that HD/H$_2$ is generally low in shocked molecular
gas while demonstrating that pressure-dependent non-LTE excitation must be
modeled explicitly.  Translating the measured HD/H$_2$ ratios into elemental
D/H requires additional modeling of shock chemistry.
\end{abstract}


\section{Introduction}

Deuterium was produced primarily during big-bang nucleosynthesis and is destroyed, rather
than created, in stars.  Thus, the deuteron is unique among the stable nuclei
in having an abundance that has decreased significantly as the Universe has aged.
Its elemental abundance probes Galactic chemical evolution together with
the infall of material of lower metallicity and higher deuterium abundance
\citep{Friedman2023}.  Precision measurements in
low-metallicity absorption systems at high-redshift give a primordial abundance
${\rm D/H}=2.53\times10^{-5}$ \citep{Cooke2018}, whereas measurements within the Galactic
disk show lower and spatially variable gas-phase abundances.  Part of that variation may
reflect astration or incomplete mixing, but preferential incorporation of D into
carbonaceous grains has also been proposed as an important cause \citep{Draine2006,Linsky2006}.
Warm shocked molecular gas offers a useful test of this picture because shocks excite the
mid-infrared rotational spectrum while potentially returning grain material to the gas.

In predominantly molecular gas, HD is the most direct tracer of deuterium.  Its
permanent dipole moment, although only $8 \times 10^{-4}$~D,
permits rotational transitions that are much faster than the
quadrupole transitions of H$_2$.  This makes HD observable as a trace species, but
also makes its excitation more difficult to interpret: HD levels can remain far from LTE
even when the corresponding H$_2$ levels are nearly thermalized.  The conversion from an
observed HD line flux to a total HD column is therefore sensitive to density as well as
temperature.  Moreover, deriving a gas-phase D/H ratio from an observed HD/H$_2$ ratio
can be complicated by the preferential destruction of HD in warm gas containing
atomic hydrogen \citep{Bertoldi1999}.

Before JWST, detections of rotationally excited interstellar HD were few and far between.
ISO detected HD toward the Orion Bar and the Orion molecular outflow, and Spitzer
detected or tentatively
detected HD in several shocked regions \citep{Neufeld2006b}.  These studies
established HD as a potentially powerful abundance and pressure diagnostic, but typically
provided only one or two HD lines per position.  JWST/MIRI now detects multiple HD lines
alongside a well-sampled H$_2$ rotational ladder.  In the JOYS survey, \citet{Francis2025},
hereafter F25, reported HD detections toward protostellar outflows across the Galactic
disk and inferred D/H abundances using LTE rotational diagrams in combination with
global corrections for non-LTE excitation and HD chemistry.

Here I develop an alternative approach in which the H$_2$ and HD rotational diagrams are
fit simultaneously with non-LTE excitation calculations.  This approach makes use of the
fact that the two molecules respond differently to gas density: H$_2$ primarily constrains
the temperature distribution and total warm-gas column, while the HD
ladder provides constraints on density and pressure.  Their relative columns then determine
$x({\rm HD})$.  Section 2 presents the excitation calculations, Section 3 describes the
joint fitting method and its application to the JOYS sample, and
Section 4 discusses the fitted physical parameters and abundances.  A brief
summary of the conclusions follows in Section 5.

\section{Non-LTE Excitation of HD and H$_2$}

\subsection{Excitation model}

I have modeled the non-LTE excitation of H$_2$ and HD rotational states over a wide range of densities
($10^{2}-10^{12}\,{\rm cm^{-3}}$) and temperatures (100--4000~K).   The calculations assume
pure collisional excitation, followed by spontaneous radiative decay,
and solve the equations of statistical equilibrium in the optically thin limit.
This limit is well justified for the applications considered here: the H$_2$ rotational
transitions are dipole-forbidden, with low radiative rates, while the relevant HD columns
and HD dipole moment are small enough that radiative trapping is negligible.

\Needspace{5\baselineskip}
\subsection{Collisional rate coefficients and spontaneous radiative decay rates}

Departures from LTE are governed by the competition between collisional excitation and
spontaneous radiative decay. For the case of H$_2$,
I adopted the collisional rate coefficients of \cite{FlowerRoueff1998} and \cite{FlowerRoueff1999a}
for the four ortho/para
H$_2$--H$_2$ combinations relevant to the present population grids.  Reactive collisions were
neglected, as were inelastic collisions in which both molecules change rotational state.
These studies provide rate coefficients for gas kinetic temperatures, $T_{\rm kin}$, spanning
the range 100--6000~K for all collisionally induced transitions among the lowest 16 rotational
states of H$_2$ in $v=0$ ($J=0$--15).
I combined the results for excitation by ortho- and para-H$_2$ colliders
assuming an ortho-to-para ratio
(OPR) of 3, the value expected in LTE at temperatures above 300~K.
Because the excitation rates are similar for ortho-H$_2$ and para-H$_2$ collision partners
(see, for example, Figure 1 of \cite{FlowerRoueff1999a}), the results are only weakly
sensitive to this assumption, as is discussed quantitatively in Appendix B.  
Excitation by He, H, or electrons was also neglected.

For HD, rate coefficients for collisional excitation by H$_2$ were adopted from
\cite{Wan2019}, who presented separate results for ortho- and para-H$_2$ over
the temperature range $\sim 10$--10000~K. Those rates
are in good agreement with earlier results obtained by \cite{FlowerRoueff1999b} for
temperatures up to 2000~K.
One significant limitation of both studies is that the calculations were
limited to states with $J_U$ up to only 8 \citep{Wan2019} or 9 \citep{FlowerRoueff1999b}.
Because JWST/MIRI can measure populations as high as $J_U = 10$
through detections of the HD R(9) transition at 12.48 $\mu$m, and because reliable modeling
requires states above those directly constrained by the observations, I
extrapolated the HD–H$_2$ rates to higher rotational states.
The extrapolation methods adopted here are described in Appendix A; they were used to 
estimate the collisional rates for
transitions between all HD rotational states up to $J_U = 14$.   Once again,
the rate coefficients for ortho- and para-H$_2$ were combined assuming an OPR of 3.

For H$_2$, the spontaneous radiative decay rates were taken from calculations of
\cite{Roueff2019}.
For HD, I used the Cologne Database for Molecular Spectroscopy values for transitions
with $J_U \le 12$ \citep{Muller2005}, which are based on the dipole-moment function of
\cite{PachuckiKomasa2008}, and extrapolated them to higher transitions.

Spontaneous radiative decay and non-reactive collisions change the spin symmetry of H$_2$ at a
negligible rate.  For the purposes of the excitation calculation, ortho- and para-H$_2$ can
therefore be treated as separate species, with uncoupled sets of statistical-equilibrium
equations (SEE).

\subsection{Steady-state level populations and rotational temperatures for transitions
observed with JWST/MIRI}

I solved the SSE on a grid with temperature points spaced by
0.01 dex and density points spaced by 0.1 dex.   Figure 1 shows representative rotational diagrams
for H$_2$ and HD at $T_{\rm kin}=1000$~K, for several gas densities.  The plotted points extend
to $J=12$ for both molecules.  The full solutions, with the population in each rotational
state given as a function of temperature and density, are provided as FITS files on Zenodo
\citep[doi:10.5281/zenodo.20663618]{NeufeldData2026}.

At high density, the rotational diagrams converge to the LTE result.  As the density is reduced,
the diagrams steepen and develop negative curvature.  In the highly subthermal regime, as shown
for HD at the lowest density in Figure 1, the curvature becomes positive.  Similar behavior is
found in isothermal excitation models for CO rotational diagrams and can be understood
analytically \citep{Neufeld2012}.

For a pair of rotational states, we may define the rotational temperature by
\[
T_{\rm rot} = {E_U-E_L \over k_B \ln[(f_L/g_L)/(f_U/g_U)]},
\]
where $E_U$ and $E_L$ are the level energies, $g_U$ and $g_L$ are the degeneracies, $f_U$ and
$f_L$ are the fractional populations, and $k_B$ is Boltzmann's constant.  Equivalently, the
local slope of a rotational diagram is
$d\ln(f_U/g_U)/dE_U = -1/(k_B T_{\rm rot})$.  When the rotational diagram is curved, the
rotational temperature therefore depends on the chosen transition or fitted range.

To define a single apparent rotational temperature for the MIRI-accessible lines, I fit a
straight line to the relevant portion of the rotational diagram and take $T_{\rm rot}$ to be
$-1/k_B$ divided by the fitted slope.  The dashed lines in Figure 1 show these fits.  For H$_2$,
I fit the upper states $3 \le J_U \le 10$, corresponding to the S(1)--S(8) transitions.  For HD,
I fit $5 \le J_U \le 10$, corresponding to the R(4)--R(9) transitions.

Figure 2 shows the resulting apparent rotational temperatures as functions of kinetic
temperature and gas density.  The left panels show $T_{\rm rot}$ versus $T_{\rm kin}$ at fixed
density, while the right panels show $T_{\rm rot}$ versus $n({\rm H_2})$ at fixed kinetic
temperature.  For both molecules, $T_{\rm rot}$ approaches $T_{\rm kin}$ in the high-density
limit.  The density required to approach LTE is about two orders of magnitude higher for HD than
for H$_2$, because the spontaneous radiative rates of the dipole-allowed HD transitions are
roughly 100 times larger than those of the H$_2$ quadrupole transitions.

Tables 1 and 2 present the grids shown in Figure 2 in tabular form, for H$_2$ and HD, respectively.  Each
table gives both the forward mapping from $(T_{\rm kin}, n)$ to the apparent MIRI rotational
temperature and the inverse mapping from an observed rotational temperature to the corresponding
kinetic temperature at a specified density.  These tables are useful for interpreting JWST H$_2$
rotational diagrams, which are often fit with two LTE temperature components.  If the gas density
is below $\sim 10^5\,{\rm cm}^{-3}$, such LTE fits can significantly underestimate the true
kinetic temperatures.  Provided that collisional excitation dominates over UV excitation and
formation pumping, an independent density estimate allows the appropriate correction to be read
from Table 1.

In the next section, I will argue that the HD rotational diagram, fitted in combination with
the H$_2$ rotational diagram, can provide such a density estimate, allowing the gas pressure to
be determined or constrained along with the HD/H$_2$ abundance ratio.

\begin{figure*}[p!]
\centering
\fullpagefig{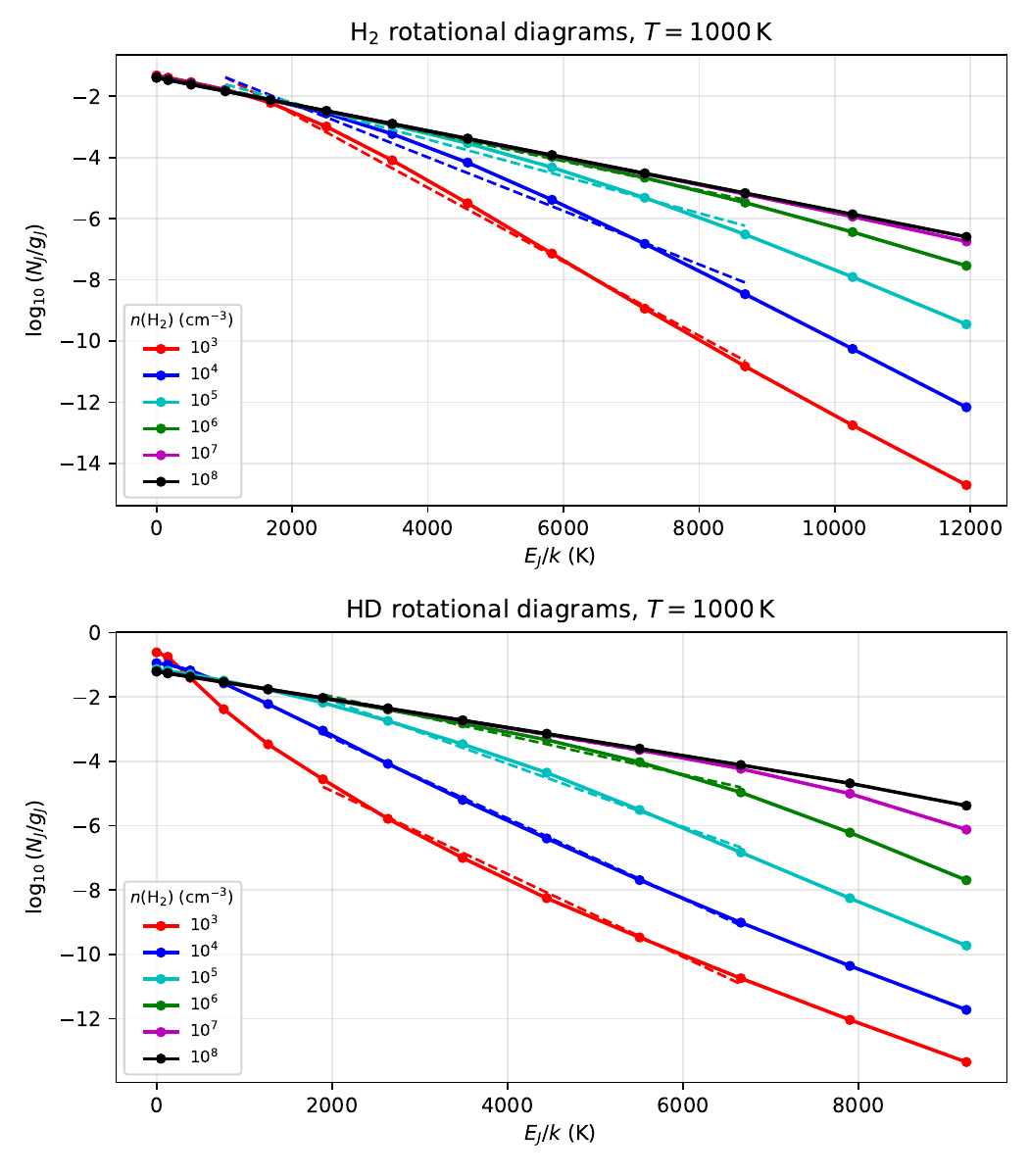}
\caption{Model rotational diagrams for H$_2$ and HD at fixed kinetic temperature,
shown for a sequence of gas densities. Dashed lines indicate the fits used to
define the apparent rotational temperatures.}
\label{fig:rot-diagrams-model}
\end{figure*}

\begin{figure*}[p!]
\centering
\fullpagefig{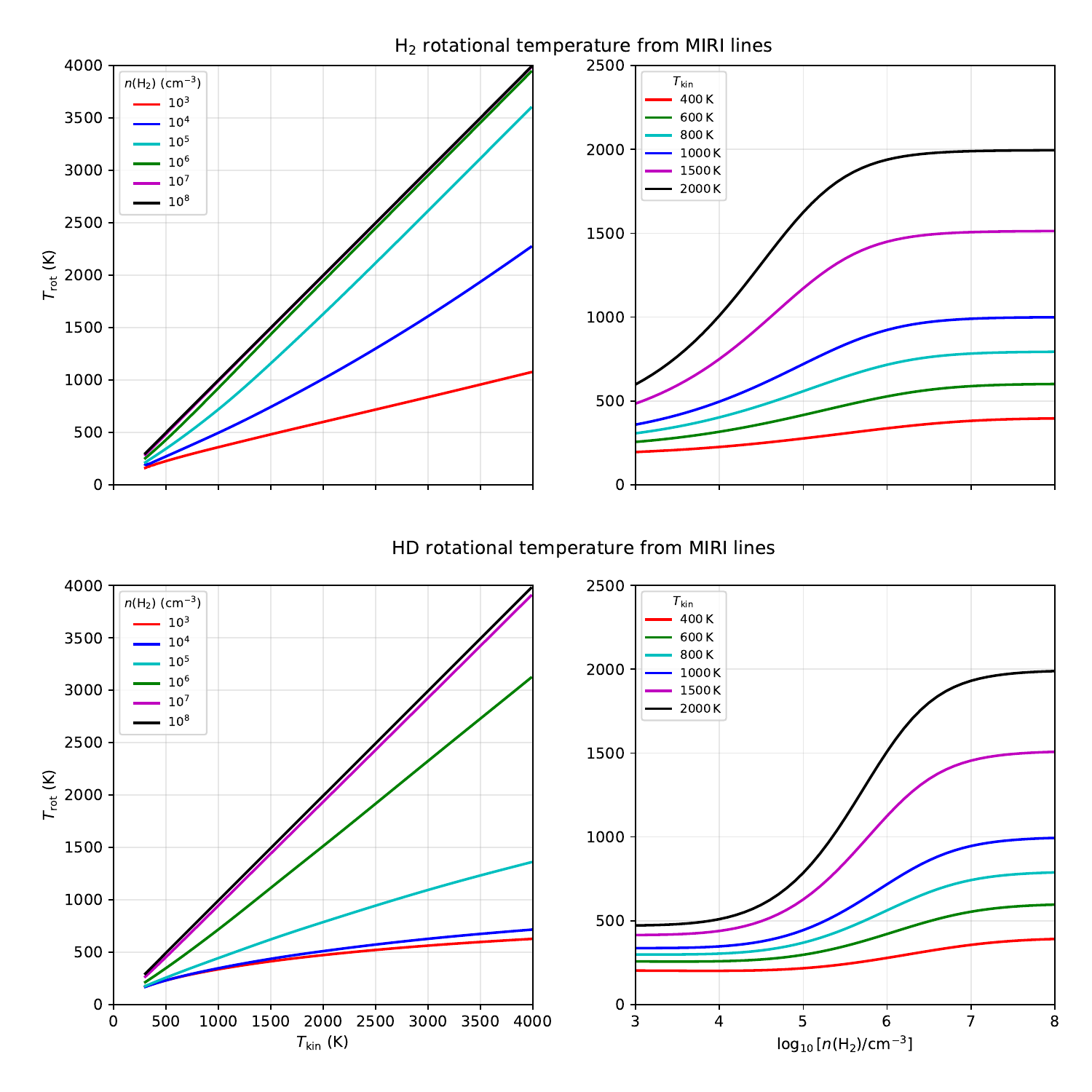}
\caption{Apparent rotational temperature as a function of kinetic temperature and
gas density. The left panels show $\trot$ versus $\tkin$ at fixed density, while the
right panels show $\trot$ versus density for fixed kinetic temperature.}
\label{fig:trot-grid}
\end{figure*}

\movetabledown=2.0in
\begin{rotatetable}
\raggedright
\refstepcounter{table}
\label{tab:h2-temperatures}
\begin{minipage}{8in}
\textbf{Table \thetable.} Conversion between fitted $T_{\rm rot}$ (MIRI lines) and $T_{\rm kin}$ for H$_2$.\par\medskip
\raggedright
\hspace*{-1.4in}%
\resizebox{\linewidth}{!}{%
\begin{tabular}{c|cccccccccc|cccccccccc}
\hline
Density & \multicolumn{10}{|c}{Forward: $T_{\rm rot}$ (K) for specified $T_{\rm kin}$ (K)} & \multicolumn{10}{|c}{Inverse: $T_{\rm kin}$ (K) for specified $T_{\rm rot}$ (K)} \\
$\log_{10} n({\rm H_2})$ & 400 & 500 & 600 & 800 & 1000 & 1200 & 1500 & 2000 & 3000 & 4000 & 400 & 500 & 600 & 800 & 1000 & 1200 & 1500 & 2000 & 3000 & 4000 \\
\hline
3.0 & 195.6 & 226.2 & 254.8 & 308.4 & 358.7 & 408.0 & 480.9 & 599.5 & 836.0 & 1078.2 & 1167.6 & 1579.7 & 2002.1 & 2848.3 & 3679.3 & NS & NS & NS & NS & NS \\
3.2 & 200.5 & 232.9 & 263.8 & 322.7 & 379.0 & 434.9 & 519.0 & 658.4 & 943.6 & 1244.7 & 1075.1 & 1432.3 & 1790.3 & 2501.0 & 3191.7 & 3854.2 & NS & NS & NS & NS \\
3.4 & 205.7 & 240.6 & 274.2 & 339.3 & 402.7 & 466.4 & 563.7 & 728.1 & 1073.1 & 1447.7 & 991.6 & 1304.2 & 1610.9 & 2214.1 & 2793.9 & 3347.6 & NS & NS & NS & NS \\
3.6 & 211.8 & 249.4 & 286.2 & 358.5 & 429.9 & 502.8 & 615.7 & 809.8 & 1226.9 & 1690.1 & 916.3 & 1192.4 & 1458.8 & 1975.3 & 2468.0 & 2938.5 & 3600.2 & NS & NS & NS \\
3.8 & 218.7 & 259.4 & 299.7 & 380.2 & 461.0 & 544.4 & 675.4 & 904.1 & 1405.8 & 1971.1 & 849.2 & 1094.6 & 1329.0 & 1775.6 & 2200.9 & 2604.0 & 3174.5 & NS & NS & NS \\
4.0 & 226.4 & 270.7 & 315.0 & 404.7 & 495.9 & 591.4 & 742.9 & 1011.1 & 1607.6 & 2282.3 & 789.6 & 1008.7 & 1217.6 & 1609.2 & 1980.0 & 2331.4 & 2829.0 & 3592.9 & NS & NS \\
4.2 & 235.1 & 283.2 & 332.0 & 431.8 & 534.8 & 643.5 & 817.9 & 1129.5 & 1825.9 & 2607.3 & 736.9 & 933.4 & 1121.4 & 1470.0 & 1796.4 & 2107.9 & 2548.3 & 3231.0 & NS & NS \\
4.4 & 244.5 & 297.0 & 350.6 & 461.5 & 577.2 & 700.5 & 899.6 & 1256.4 & 2049.6 & 2922.7 & 690.2 & 867.6 & 1038.0 & 1352.8 & 1644.5 & 1923.4 & 2320.6 & 2940.7 & NS & NS \\
4.6 & 254.8 & 311.9 & 370.6 & 493.4 & 622.8 & 761.4 & 985.8 & 1386.5 & 2264.0 & 3205.0 & 648.9 & 810.4 & 965.6 & 1253.3 & 1518.4 & 1771.3 & 2135.5 & 2708.6 & 3784.8 & NS \\
4.8 & 265.8 & 327.7 & 391.9 & 527.2 & 670.7 & 824.8 & 1073.5 & 1513.3 & 2455.0 & 3437.8 & 612.3 & 760.7 & 903.0 & 1168.8 & 1413.4 & 1646.6 & 1985.2 & 2524.9 & 3556.6 & NS \\
5.0 & 277.4 & 344.4 & 414.2 & 562.2 & 719.7 & 888.5 & 1158.7 & 1629.5 & 2612.7 & 3616.1 & 580.0 & 717.5 & 849.1 & 1096.8 & 1326.0 & 1544.7 & 1864.2 & 2381.3 & 3386.5 & NS \\
5.2 & 289.5 & 361.6 & 437.2 & 597.7 & 768.3 & 950.0 & 1237.2 & 1729.3 & 2734.2 & 3744.9 & 551.4 & 679.9 & 802.8 & 1035.8 & 1253.3 & 1461.7 & 1768.0 & 2271.1 & 3262.8 & NS \\
5.4 & 301.8 & 379.1 & 460.3 & 632.5 & 814.6 & 1006.5 & 1305.5 & 1809.4 & 2822.4 & 3833.9 & 526.2 & 647.4 & 763.2 & 984.3 & 1193.4 & 1395.0 & 1693.1 & 2188.4 & 3175.3 & NS \\
5.6 & 314.2 & 396.5 & 482.9 & 665.5 & 856.7 & 1055.8 & 1361.2 & 1870.0 & 2883.7 & 3893.6 & 504.1 & 619.2 & 729.6 & 941.5 & 1144.6 & 1342.1 & 1636.0 & 2128.0 & 3115.0 & NS \\
5.8 & 326.5 & 413.5 & 504.5 & 695.4 & 893.3 & 1096.3 & 1404.2 & 1913.4 & 2924.9 & 3933.0 & 484.8 & 595.1 & 701.2 & 906.4 & 1105.6 & 1301.1 & 1593.6 & 2085.3 & 3074.5 & NS \\
6.0 & 338.4 & 429.5 & 524.4 & 721.5 & 923.2 & 1127.9 & 1435.6 & 1943.4 & 2952.0 & 3958.6 & 468.1 & 574.6 & 677.5 & 878.2 & 1075.2 & 1270.2 & 1563.1 & 2056.0 & 3047.7 & NS \\
6.2 & 349.6 & 444.2 & 542.1 & 743.1 & 946.5 & 1151.2 & 1457.6 & 1963.4 & 2969.6 & 3975.2 & 453.8 & 557.4 & 658.1 & 856.1 & 1052.3 & 1247.6 & 1541.7 & 2036.3 & 3030.3 & NS \\
6.4 & 359.9 & 457.2 & 557.0 & 760.0 & 963.8 & 1167.7 & 1472.5 & 1976.6 & 2980.9 & 3985.8 & 441.6 & 543.1 & 642.5 & 839.2 & 1035.5 & 1231.7 & 1527.2 & 2023.3 & 3019.0 & NS \\
6.6 & 369.0 & 468.1 & 569.2 & 772.7 & 976.0 & 1179.0 & 1482.3 & 1985.1 & 2988.1 & 3992.5 & 431.6 & 531.7 & 630.3 & 826.8 & 1023.6 & 1220.8 & 1517.6 & 2014.9 & 3011.8 & NS \\
6.8 & 376.7 & 477.1 & 578.5 & 781.8 & 984.3 & 1186.4 & 1488.7 & 1990.5 & 2992.7 & 3996.8 & 423.4 & 522.7 & 621.1 & 817.9 & 1015.5 & 1213.4 & 1511.2 & 2009.4 & 3007.3 & NS \\
7.0 & 383.1 & 484.0 & 585.4 & 788.1 & 989.9 & 1191.3 & 1492.8 & 1994.0 & 2995.6 & 3999.5 & 416.8 & 515.8 & 614.3 & 811.8 & 1010.0 & 1208.6 & 1507.2 & 2006.0 & 3004.4 & NS \\
7.2 & 388.1 & 489.1 & 590.3 & 792.3 & 993.5 & 1194.5 & 1495.4 & 1996.2 & 2997.5 & 4001.3 & 411.8 & 510.8 & 609.5 & 807.7 & 1006.4 & 1205.5 & 1504.6 & 2003.8 & 3002.5 & 3998.8 \\
7.4 & 391.8 & 492.8 & 593.7 & 795.1 & 995.9 & 1196.5 & 1497.1 & 1997.6 & 2998.7 & 4002.3 & 408.1 & 507.2 & 606.3 & 804.9 & 1004.1 & 1203.5 & 1502.9 & 2002.4 & 3001.3 & 3997.7 \\
7.6 & 394.5 & 495.3 & 595.9 & 796.8 & 997.4 & 1197.8 & 1498.2 & 1998.5 & 2999.4 & 4003.0 & 405.4 & 504.7 & 604.0 & 803.1 & 1002.6 & 1202.2 & 1501.8 & 2001.5 & 3000.6 & 3997.0 \\
7.8 & 396.4 & 496.9 & 597.4 & 798.0 & 998.3 & 1198.6 & 1498.8 & 1999.1 & 2999.9 & 4003.5 & 403.6 & 503.1 & 602.6 & 802.0 & 1001.7 & 1201.4 & 1501.2 & 2000.9 & 3000.1 & 3996.6 \\
8.0 & 397.7 & 498.0 & 598.3 & 798.7 & 999.0 & 1199.1 & 1499.3 & 1999.4 & 3000.2 & 4003.7 & 402.3 & 502.0 & 601.7 & 801.3 & 1001.0 & 1200.9 & 1500.7 & 2000.6 & 2999.8 & 3996.3 \\
\hline
\end{tabular}%
}
\par\smallskip{\scriptsize \textsc{Note}---Density is in units of cm$^{-3}$. NS indicates that no solution exists over the modeled temperature range.}
\end{minipage}
\end{rotatetable}
\thispagestyle{empty}
\clearpage

\movetabledown=2.0in
\begin{rotatetable}
\raggedright
\refstepcounter{table}
\label{tab:hd-temperatures}
\begin{minipage}{8in}
\textbf{Table \thetable.} Conversion between fitted $T_{\rm rot}$ (MIRI lines) and $T_{\rm kin}$ for HD.\par\medskip
\raggedright
\hspace*{-1.4in}%
\resizebox{\linewidth}{!}{%
\begin{tabular}{c|cccccccccc|cccccccccc}
\hline
Density & \multicolumn{10}{|c}{Forward: $T_{\rm rot}$ (K) for specified $T_{\rm kin}$ (K)} & \multicolumn{10}{|c}{Inverse: $T_{\rm kin}$ (K) for specified $T_{\rm rot}$ (K)} \\
$\log_{10} n({\rm H_2})$ & 400 & 500 & 600 & 800 & 1000 & 1200 & 1500 & 2000 & 3000 & 4000 & 400 & 500 & 600 & 800 & 1000 & 1200 & 1500 & 2000 & 3000 & 4000 \\
\hline
3.0 & 203.7 & 232.1 & 257.1 & 300.3 & 337.3 & 370.1 & 413.3 & 472.8 & 563.1 & 627.8 & 1402.5 & 2266.8 & 3533.8 & NS & NS & NS & NS & NS & NS & NS \\
3.2 & 203.2 & 231.6 & 256.6 & 300.1 & 337.4 & 370.5 & 414.2 & 474.5 & 566.5 & 632.5 & 1397.4 & 2244.4 & 3471.0 & NS & NS & NS & NS & NS & NS & NS \\
3.4 & 202.7 & 231.2 & 256.3 & 300.1 & 337.9 & 371.6 & 416.2 & 477.9 & 572.6 & 640.8 & 1385.8 & 2205.2 & 3367.6 & NS & NS & NS & NS & NS & NS & NS \\
3.6 & 202.4 & 230.9 & 256.3 & 300.8 & 339.3 & 373.9 & 420.0 & 484.0 & 583.0 & 655.2 & 1364.5 & 2141.6 & 3210.5 & NS & NS & NS & NS & NS & NS & NS \\
3.8 & 202.1 & 231.0 & 256.8 & 302.4 & 342.2 & 378.2 & 426.6 & 494.2 & 600.3 & 678.9 & 1330.1 & 2047.8 & 2996.2 & NS & NS & NS & NS & NS & NS & NS \\
4.0 & 202.2 & 231.7 & 258.2 & 305.5 & 347.3 & 385.5 & 437.3 & 510.5 & 627.9 & 716.5 & 1280.2 & 1922.1 & 2735.2 & NS & NS & NS & NS & NS & NS & NS \\
4.2 & 202.9 & 233.2 & 260.9 & 310.8 & 355.6 & 397.1 & 454.0 & 535.7 & 669.9 & 773.8 & 1214.8 & 1770.7 & 2448.6 & NS & NS & NS & NS & NS & NS & NS \\
4.4 & 204.5 & 236.2 & 265.4 & 319.2 & 368.2 & 414.3 & 478.6 & 572.6 & 731.5 & 858.1 & 1136.7 & 1606.7 & 2158.9 & 3515.5 & NS & NS & NS & NS & NS & NS \\
4.6 & 207.2 & 240.9 & 272.5 & 331.5 & 386.5 & 439.0 & 513.5 & 624.6 & 818.7 & 978.0 & 1050.9 & 1443.7 & 1883.6 & 2895.1 & NS & NS & NS & NS & NS & NS \\
4.8 & 211.6 & 248.0 & 282.6 & 348.7 & 411.5 & 472.6 & 560.9 & 695.3 & 937.6 & 1143.1 & 962.8 & 1291.4 & 1639.6 & 2415.8 & 3286.5 & NS & NS & NS & NS & NS \\
5.0 & 217.9 & 257.8 & 296.4 & 371.5 & 444.3 & 516.5 & 622.5 & 787.4 & 1094.2 & 1362.9 & 877.6 & 1154.2 & 1435.2 & 2039.4 & 2681.3 & 3378.2 & NS & NS & NS & NS \\
5.2 & 226.3 & 270.5 & 314.0 & 400.4 & 485.6 & 571.4 & 699.4 & 902.7 & 1291.8 & 1644.2 & 799.2 & 1033.7 & 1266.6 & 1743.2 & 2245.1 & 2757.0 & 3579.7 & NS & NS & NS \\
5.4 & 236.8 & 286.3 & 335.7 & 435.2 & 535.2 & 637.1 & 791.1 & 1040.1 & 1528.8 & 1986.0 & 729.4 & 929.8 & 1127.7 & 1517.4 & 1918.2 & 2324.3 & 2939.5 & NS & NS & NS \\
5.6 & 249.5 & 305.0 & 361.0 & 475.5 & 592.1 & 711.9 & 894.9 & 1194.8 & 1795.0 & 2372.5 & 668.7 & 842.3 & 1013.4 & 1344.5 & 1673.6 & 2008.7 & 2507.3 & 3350.2 & NS & NS \\
5.8 & 264.0 & 326.1 & 389.3 & 519.9 & 654.1 & 792.7 & 1005.2 & 1357.0 & 2069.3 & 2769.9 & 616.7 & 769.9 & 920.0 & 1210.3 & 1492.6 & 1776.0 & 2202.3 & 2902.4 & NS & NS \\
6.0 & 279.9 & 348.9 & 419.6 & 566.5 & 718.0 & 874.6 & 1114.5 & 1513.4 & 2324.8 & 3134.0 & 572.5 & 710.4 & 844.7 & 1105.5 & 1357.2 & 1606.8 & 1983.3 & 2601.1 & 3835.0 & NS \\
6.2 & 296.7 & 372.5 & 450.6 & 612.8 & 779.9 & 951.8 & 1214.2 & 1650.9 & 2538.2 & 3428.8 & 535.6 & 661.9 & 784.5 & 1023.7 & 1255.3 & 1483.8 & 1827.0 & 2395.6 & 3519.4 & NS \\
6.4 & 313.5 & 395.8 & 480.6 & 655.9 & 835.6 & 1019.2 & 1298.0 & 1761.2 & 2698.6 & 3641.4 & 505.0 & 622.5 & 736.9 & 960.7 & 1179.2 & 1394.7 & 1717.8 & 2256.3 & 3320.4 & NS \\
6.6 & 329.8 & 417.9 & 508.2 & 693.7 & 882.4 & 1073.8 & 1362.8 & 1842.5 & 2809.3 & 3781.3 & 479.9 & 591.0 & 699.6 & 913.0 & 1123.2 & 1331.0 & 1642.8 & 2163.7 & 3196.6 & NS \\
6.8 & 344.9 & 437.6 & 532.1 & 724.6 & 919.0 & 1114.9 & 1409.6 & 1898.6 & 2881.2 & 3868.0 & 459.7 & 566.3 & 671.0 & 877.8 & 1082.9 & 1286.6 & 1592.3 & 2103.5 & 3120.6 & NS \\
7.0 & 358.3 & 454.3 & 551.6 & 748.3 & 945.8 & 1144.1 & 1441.6 & 1935.6 & 2926.4 & 3920.3 & 443.7 & 547.1 & 649.4 & 852.4 & 1054.7 & 1256.3 & 1559.0 & 2065.1 & 3074.2 & NS \\
7.2 & 369.5 & 467.6 & 566.6 & 765.5 & 964.6 & 1163.9 & 1462.7 & 1959.4 & 2954.3 & 3951.6 & 431.2 & 532.8 & 633.7 & 834.6 & 1035.6 & 1236.2 & 1537.5 & 2040.9 & 3045.9 & NS \\
7.4 & 378.4 & 477.8 & 577.6 & 777.5 & 977.2 & 1176.9 & 1476.4 & 1974.4 & 2971.6 & 3970.5 & 421.8 & 522.3 & 622.5 & 822.6 & 1022.9 & 1223.1 & 1523.7 & 2025.7 & 3028.5 & NS \\
7.6 & 385.1 & 485.1 & 585.2 & 785.5 & 985.4 & 1185.4 & 1485.1 & 1983.9 & 2982.3 & 3981.9 & 414.9 & 514.8 & 614.7 & 814.5 & 1014.6 & 1214.7 & 1515.0 & 2016.1 & 3017.7 & NS \\
7.8 & 390.0 & 490.2 & 590.4 & 790.7 & 990.7 & 1190.7 & 1490.6 & 1989.9 & 2988.9 & 3988.9 & 409.9 & 509.7 & 609.5 & 809.3 & 1009.3 & 1209.3 & 1509.4 & 2010.2 & 3011.1 & NS \\
8.0 & 393.4 & 493.7 & 593.9 & 794.1 & 994.1 & 1194.1 & 1494.0 & 1993.6 & 2993.1 & 3993.1 & 406.5 & 506.3 & 606.1 & 805.9 & 1005.9 & 1205.9 & 1506.0 & 2006.4 & 3006.9 & NS \\
\hline
\end{tabular}%
}
\par\smallskip{\scriptsize \textsc{Note}---Density is in units of cm$^{-3}$. NS indicates that no solution exists over the modeled temperature range.}
\end{minipage}
\end{rotatetable}
\thispagestyle{empty}
\clearpage

\section{Fitting Rotational Diagrams obtained from JWST}

\subsection{Two-component fits to H$_2$ rotational diagrams}

The Infrared Spectrograph (IRS) on Spitzer provided
an extensive collection of H$_2$ rotational diagrams, obtained
from observations of the S(0) through S(7) pure rotational lines
\citep[e.g.,][]{Neufeld2006a,Neufeld2007,Giannini2011}
carried out toward warm, shock-heated interstellar gas, along with
a few detections of the HD R(3) and R(4) lines
\citep{Neufeld2006b,Neufeld2007,Giannini2011}.  The H$_2$ rotational
diagrams obtained in these studies typically show positive curvature, in the sense
that high-$J$ levels lie above the extrapolation from the low-$J$ levels.  Because
subthermal excitation does not produce such positive curvature in H$_2$ rotational
diagrams over the density range relevant to these observations (see Figure 1, for example), the observed
curvature is strong evidence that a range of gas temperatures is present
within the beam, resulting perhaps from a range of shock velocities
\citep[e.g.,][]{Neufeld2006a,NeufeldYuan2008,Giannini2011}.  The data were
typically fit with a two-component model, in which the emission was assumed to arise
from a warm gas component in LTE at a temperature typically below $1000$~K and a hot component in
LTE at a higher temperature, or alternatively with a model in which a continuous power-law
distribution of temperatures was posited \citep[e.g.,][]{NeufeldYuan2008,Giannini2011}.  While the latter approach is arguably more
realistic, it does not typically fit the data any better.

In most instances, the H$_2$ rotational diagrams obtained from Spitzer also showed
the ``zig-zag'' behavior characteristic of a non-equilibrium OPR (i.e. with
even-$J$ para-states elevated relative to the adjacent odd-$J$ states), suggesting
that the gas retained the memory of an earlier period in which it had been cooler
and possessed a lower OPR \citep[e.g.,][]{Neufeld2006a,Giannini2011}.  Moreover, the degree of ``zig-zag'' was a decreasing function of
excitation energy, prompting many Spitzer studies to fit H$_2$ rotational diagrams with
two gas components, each with its own column density, temperature and OPR.  This method
introduces six free parameters: $N_w$, $T_w$, ${\rm OPR}_w$, $N_h$, $T_h$, and ${\rm OPR}_h$, where
$N$ is the H$_2$ column density and the subscripts $w$ and $h$ denote the warm and hot
components.  With eight measured column densities (for the $J_U = 2-9$ rotational
states), the six-parameter model retains two degrees of freedom and can generally be
constrained by the observations.

Relative to Spitzer, JWST provides greatly improved spatial and spectral resolution,
particularly in the spectral region below 10~$\mu$m.  While Spitzer lacked the
spectral resolution to permit the detection of HD lines higher than R(4), the R(5)--R(9)
lines have been detected in multiple sources with JWST/MIRI by F25.  In addition, the spectroscopic
capabilities of JWST extend to shorter wavelengths, allowing the H$_2$ S(8) line to be detected
with the MIRI instrument and pure rotational lines as high as S(17) to be detected with
NIRSpec \citep[e.g.,][]{Federman2024,Narang2024}; larger samples are expected from programs
such as HEFE \citep{Megeath2024}.  (On the other hand, the H$_2$ S(0) and HD R(3) lines at 28.22 and 28.50~$\mu$m
lie at wavelengths close to the long-wavelength cutoff for MIRI, where the sensitivity
is dropping rapidly, and are not typically detectable.)

With JWST's improved spatial resolution and sensitivity, H$_2$ lines
have now been detected in deeply embedded gas where dust extinction is very
significant even at mid-IR wavelengths \citep[e.g.,][]{Federman2024,Narang2024,Tyagi2025}.  Several groups have added extinction as
a seventh free parameter in fitting the H$_2$ rotational diagrams, usually adopting the
KP5 extinction curve \citep{Pontoppidan2024} to apply an extinction correction to all the measured line fluxes
\citep[e.g.,][]{Francis2025,Neufeld2024}.
Because the S(3) line at 9.66~$\mu$m lies very near the peak of the 9.7~$\mu$m
silicate absorption feature, it is the most strongly extinguished line and
provides good leverage for determining the extinction to the emitting gas.  This also
makes the inferred extinction somewhat sensitive to the assumed extinction curve near
the silicate feature.  Rather than
simply being a ``nuisance parameter'' in the fitting of H$_2$ rotational diagrams,
extinction estimates from H$_2$ have been used to deredden spectral lines of
other species.  Whereas the traditional two-component fits described above assume LTE,
the simultaneous H$_2$--HD fits described below calculate the excitation of both species
in non-LTE.

\Needspace{5\baselineskip}
\subsection{Fitting the H$_2$ and HD rotational diagrams simultaneously}

In cases where emission lines from HD have been detected along with those from H$_2$,
the two rotational diagrams can be fitted simultaneously to constrain the gas pressure and
the HD/H$_2$ abundance ratio.  The H$_2$ diagram constrains the column densities,
temperatures, and OPRs of the warm and hot components, while the HD diagram adds a
complementary constraint on the density.  This is because the HD rotational transitions have much
larger spontaneous radiative decay rates than the corresponding H$_2$ quadrupole transitions,
and the HD levels therefore remain subthermally excited to higher densities.  When the HD
populations approach LTE, however, the rotational diagram becomes insensitive to further
increases in density and may provide only a lower limit on the gas pressure.

The fitting model used below keeps the same two gas components used for the H$_2$ rotational
diagram, but requires the H$_2$ and HD emission to arise from gas with the same component
temperatures and, within each component, identical beam-filling factors.  The two components
are assumed to be isobaric and are therefore described by a common gas pressure,
$p$, so that their densities are set by
their fitted kinetic temperatures.  A single abundance
parameter,
\[
x({\rm HD}) = {N({\rm HD}) \over N({\rm H_2})},
\]
is also assumed to apply to both components and scales the HD columns relative to the
corresponding H$_2$ columns.  These common-pressure, common-abundance, and common-beam-filling
assumptions reduce the number of parameters and allow the complementary excitation properties
of H$_2$ and HD to constrain the model.

I have also considered separately the isochoric case, in which the densities of the
two components are assumed equal.  When applied to the dataset discussed in Section 3.3
below, the derived values of $x({\rm HD})$ are consistent within the uncertainties
with those obtained in the isobaric case (see Figure 13 in Appendix B).  The derived
densities generally lie between the values $p/T_w$ and $p/T_h$ obtained for the warm
and hot components in the isobaric case.  I also considered a model in which
$x({\rm HD})$ is allowed to differ in the hot and warm components, but this yielded
poorly constrained values for the HD abundance in the hot component: the data obtained
to date do not meaningfully constrain a temperature-dependent HD abundance.

For any trial pressure and set of component temperatures, column densities, and OPRs, the non-LTE calculations described
in Section 2 give the predicted level populations for each molecule; the model rotational
diagram is the sum of the warm and hot component contributions.
The excitation calculations assume an OPR of 3 for the H$_2$ collision partners, even when
the fitted OPR of an emitting component differs from 3; as discussed in Section 2.2, the
excitation rates are only weakly sensitive to the collider OPR.  A single extinction is fitted
jointly with the other parameters and applied consistently to both species using the same
extinction curve.

As described in detail in Appendix \ref{app:fitting-details}, two methods were
used to estimate the nine free parameters:
$N_w$, $T_w$, ${\rm OPR}_w$, $N_h$, $T_h$, ${\rm OPR}_h$, $p/k_B$,
$x({\rm HD})$, and $\tau_{\rm 9.7}$, where $\tau_{\rm 9.7}$ is the dust extinction
optical depth at 9.7~$\mu$m (i.e., at the peak of the silicate feature).
The first method uses a bounded trust-region reflective least-squares algorithm to determine
the model parameters that minimize $\chi^2$, with bounds on all fitted parameters.  This approach proves
computationally inexpensive, and also provides estimates of the parameter uncertainties as
well as the best-fit parameters.  However, the parameter uncertainty estimates obtained with
this implementation are only approximate.  In particular, the formal uncertainty in
$\log_{10}(p/k_B)$ becomes unreliable near LTE, where the likelihood is insensitive to
further increases in pressure and the pressure bound is reached for some apertures.  A second,
more computationally expensive method samples the nine-dimensional parameter space using the
Markov chain Monte Carlo (MCMC) proposed by \cite{GoodmanWeare2010} and implemented by
\cite{ForemanMackey2013}.  Given flat Bayesian priors with hard bounds
$400 \le T_w \le 1000$~K, $1000 \le T_h \le 3900$~K, $16 \le \log_{10}(N_w/{\rm cm}^{-2}) \le 22$,
$12 \le \log_{10}(N_h/{\rm cm}^{-2}) \le 21$, $0.01 \le {\rm OPR}_w \le 5$, $0.01 \le {\rm OPR}_h \le 5$,
$0 \le \tau_{\rm 9.7} \le 10$, $0 \le x({\rm HD}) \le 10^{-4}$, and $6 \le \log_{10}[p/k_B\,({\rm K\,cm^{-3}})] \le 10$,
this method yields posterior probability density functions (PDFs) for each parameter, without
assuming Gaussianity, and joint 2D PDFs for each pair of parameters; these may be
conveniently plotted in a so-called ``corner plot.''

\subsection{Application to the JOYS sample of HD detections}

As an application of the H$_2$-HD fitting method described above, I have fitted the
H$_2$ and HD fluxes reported by F25.  These were measured with MIRI toward 26 positions
in 10 protostellar outflows that were observed in the
``JWST Observations of Young protoStars'' (JOYS) program (P.I., E.\ F.\ van Dishoeck)
and were presented in Tables F.1 and F.2 of F25.  At 17 of the 26 positions observed,
HD was securely detected in at least three lines; 16 of these positions had detections of
R(4), R(5), and R(6), while BHR71-IRS1 Aperture 3 had detections of R(5)--R(9).
The H$_2$ lines were much stronger than those of HD, allowing the detection of every line
accessible to MIRI, S(1)--S(8), at every position except for an upper limit on S(6) toward
G28-IRS2 Aperture 1.  I added a 6\% fractional uncertainty in
quadrature to each line flux; this is the absolute
spectrophotometric calibration uncertainty estimated for the
MIRI/MRS from in-flight performance measurements by \citet{argyriou2023}.

Figure 3 shows the fits to the rotational diagrams obtained at two example positions:
BHR71 Aperture 2 and HH 211 Aperture 7.   As usual, the vertical axis shows $\log_{10}
(N_J/g_J)$ as a function of $E_J/k_B$, where $N_J$ is the column density in state $J$ and
$g_J$ is the degeneracy; for H$_2$, this includes the nuclear spin degeneracy factor
(3 for odd-$J$ ortho-states and 1 for even-$J$ para-states).
Black points indicate the column densities
inferred without applying an extinction correction to the fluxes.  Red, green and blue
points indicate the para-H$_2$, ortho-H$_2$, and HD column densities obtained from the
dereddened fluxes.  The red, green and blue curves show the best fits from the model.
The inset box lists the values for each of the model parameters (median with central 68\%
credible intervals) and the best-fit reduced $\chi^2$.  In both example cases, the model
provides an excellent fit to the observed column densities.  The corresponding corner
plots are shown in Figures 4 and 5.  The derived HD abundances and pressures for the
17 HD detections are given in Table 3, along with HD abundance upper limits for the
remaining nine apertures.  For the detections, the best-fit values, median values and the
central 68\% credible intervals are listed.  The quoted credible intervals reflect statistical
uncertainties only and do not include systematic uncertainties associated with the collisional
rate coefficients, extinction law, or assumptions of the excitation model.
As discussed in Appendix \ref{app:fitting-details},
the best-fit values (i.e., the values that minimize
$\chi^2$) may differ from the median values (i.e., the values above and below which
each parameter has equal posterior probability) because the posterior
distributions can be asymmetric and reflect parameter-volume effects.
In the next section, I will discuss the results obtained
from this application of the H$_2$-HD fitting method.

\begin{figure*}[p!]
\centering
\fullpagefig{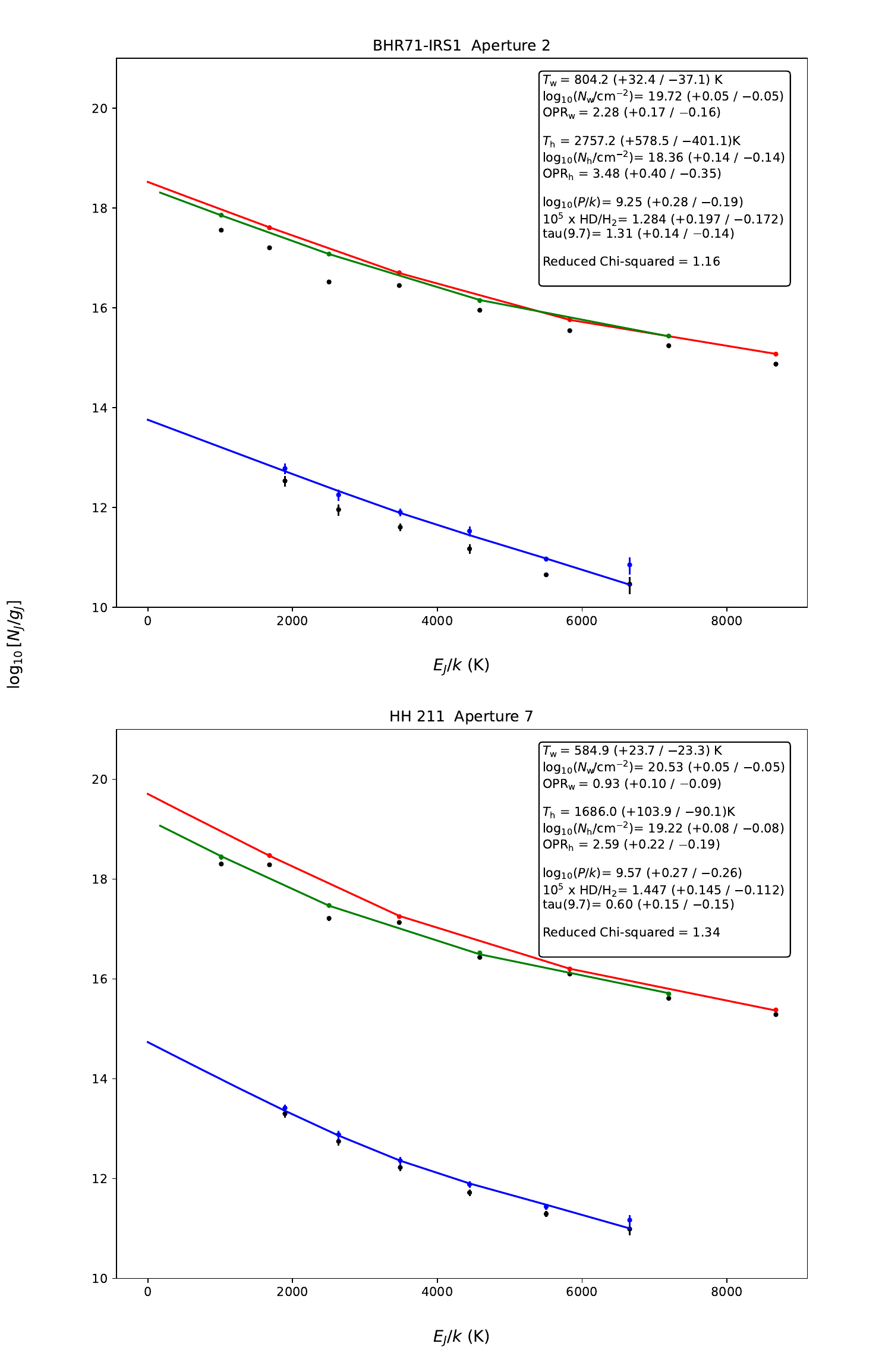}
\caption{Example rotational diagrams for BHR71 Aperture 2 and HH 211 Aperture 7.
Black points indicate the column densities
inferred without applying an extinction correction to the fluxes.  Red, green and blue
points indicate the para-H$_2$, ortho-H$_2$, and HD column densities obtained from the
dereddened fluxes.  The red, green and blue curves show the best fits from the model.}
\label{fig:example-rotdiag-fits}
\end{figure*}

\begin{figure*}[p!]
\centering
\fullpagefig{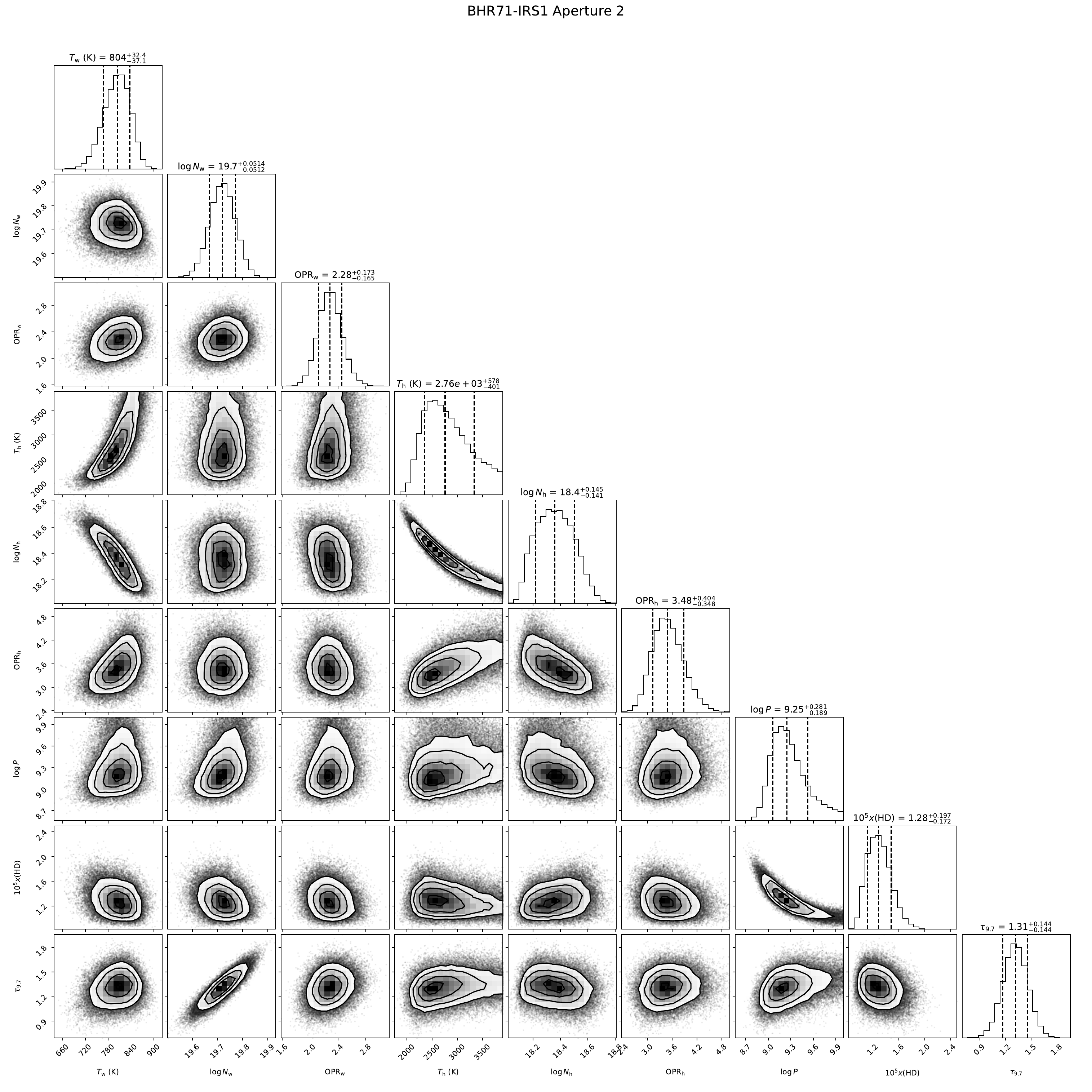}
\caption{Corner plot for the BHR71 Aperture 2 fit.}
\label{fig:corner-bhr71-ap2}
\end{figure*}

\begin{figure*}[p!]
\centering
\fullpagefig{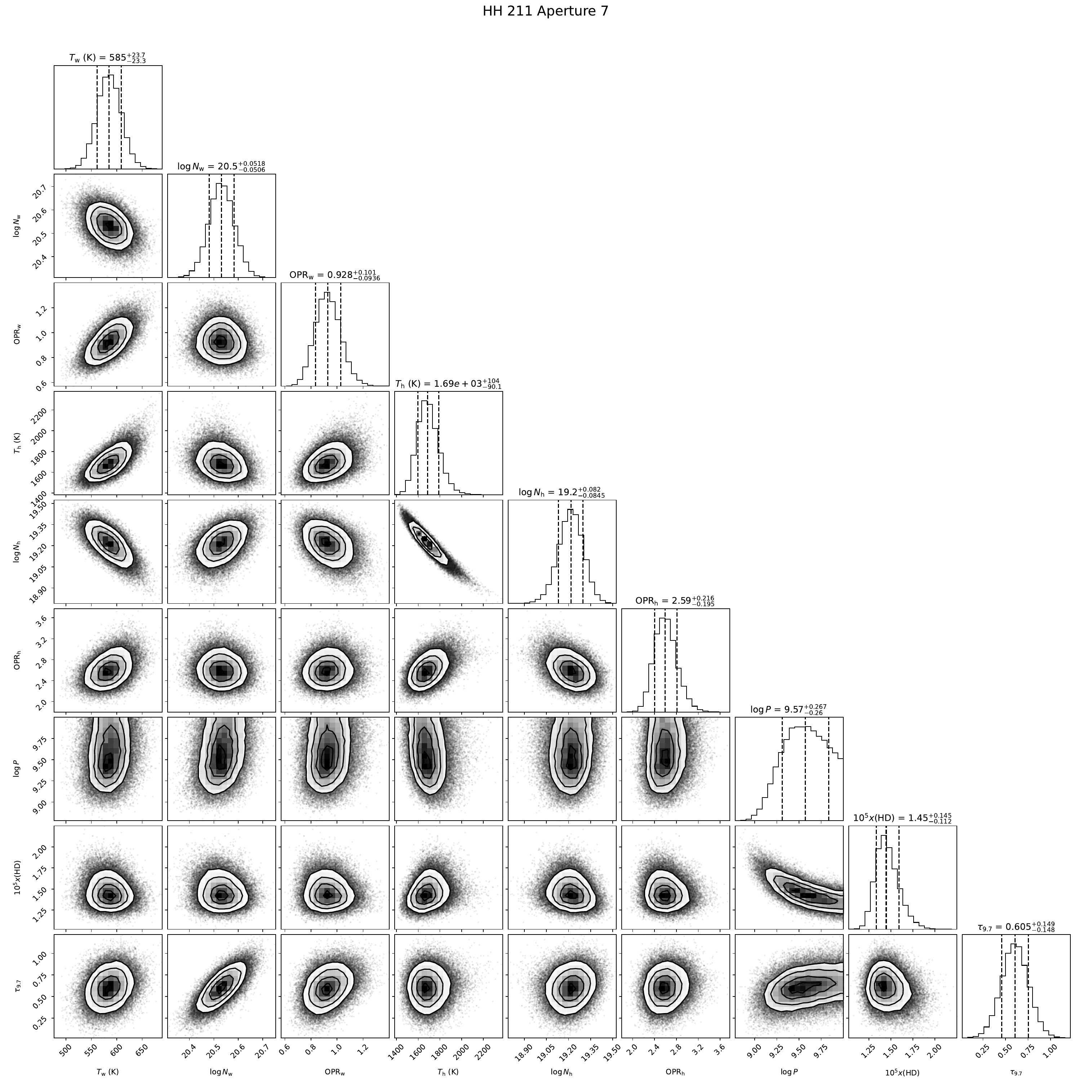}
\caption{Corner plot for the HH 211 Aperture 7 fit.}
\label{fig:corner-hh211-ap7}
\end{figure*}

\begin{table*}[p!]
\centering
\scriptsize
\setlength{\tabcolsep}{2.0pt}
\caption{Summary of fitted HD abundance and pressure. The abundance columns report $10^5 x({\rm HD})$. The credible intervals are the central 68\% MCMC intervals. Rows without HD detections list the MCMC 84th percentile as an upper limit and leave the remaining entries blank. The pressure columns report $\log_{10}[p/k_B\,({\rm K\,cm^{-3}})]$. The final column gives the reduced $\chi^2$ of the best fit.}
\label{tab:fit-results}
\begin{tabular}{l|c c c|c c c|c}
\hline
Source/aperture & \multicolumn{3}{c|}{$10^5 x({\rm HD})$} & \multicolumn{3}{c|}{$\log_{10}(p/k_B)$} & Reduced $\chi^2$ \\
 & Median & 68\% credible range & Best fit & Median & 68\% credible range & Best fit & \\
\hline
G28-IRS2 Aperture 1 &  & $<$ 6.15 &  &  &  &  &  \\
G28-P1-A Aperture 1 &  & $<$ 3.47 &  &  &  &  &  \\
G31-A Aperture 1 &  & $<$ 7.81 &  &  &  &  &  \\
IRAS 18089-1732 Aperture 1 &  & $<$ 3.00 &  &  &  &  &  \\
IRAS 23385+6053 Aperture 1 & 2.57 & 2.40--2.77 & 2.49 & 9.75 & 9.49--9.93 & 10.00 & 2.25 \\
BHR71-IRS1 Aperture 1 & 1.65 & 1.55--1.78 & 1.60 & 9.73 & 9.45--9.92 & 10.00 & 2.46 \\
BHR71-IRS1 Aperture 2 & 1.28 & 1.11--1.48 & 1.36 & 9.25 & 9.06--9.53 & 9.16 & 1.16 \\
BHR71-IRS1 Aperture 3 & 1.10 & 0.97--1.24 & 1.13 & 9.30 & 9.11--9.58 & 9.24 & 2.31 \\
BHR71-IRS1 Aperture 4 &  & $<$ 2.55 &  &  &  &  &  \\
HH 211 Aperture 1 & 0.99 & 0.89--1.11 & 0.97 & 9.51 & 9.11--9.85 & 9.58 & 1.32 \\
HH 211 Aperture 2 & 1.09 & 0.76--1.66 & 1.11 & 8.47 & 8.08--9.17 & 8.44 & 0.24 \\
HH 211 Aperture 3 & 1.06 & 0.84--1.41 & 1.08 & 8.56 & 8.24--9.04 & 8.52 & 0.39 \\
HH 211 Aperture 4 &  & $<$ 1.10 &  &  &  &  &  \\
HH 211 Aperture 5 &  & $<$ 1.35 &  &  &  &  &  \\
HH 211 Aperture 6 &  & $<$ 6.36 &  &  &  &  &  \\
HH 211 Aperture 7 & 1.45 & 1.34--1.59 & 1.45 & 9.57 & 9.31--9.84 & 9.53 & 1.34 \\
HH 211 Aperture 8 & 1.42 & 1.33--1.51 & 1.40 & 9.90 & 9.77--9.97 & 10.00 & 2.13 \\
HH 211 Aperture 9 & 2.01 & 1.82--2.23 & 1.88 & 9.71 & 9.45--9.91 & 9.96 & 1.79 \\
IRAS 4B Aperture 1 & 3.58 & 3.33--3.84 & 3.55 & 9.92 & 9.82--9.98 & 10.00 & 5.87 \\
IRAS 4B Aperture 2 & 3.85 & 3.55--4.15 & 3.81 & 9.91 & 9.80--9.98 & 10.00 & 6.92 \\
L1448-mm Aperture 1 & 1.68 & 1.58--1.79 & 1.65 & 9.70 & 9.44--9.91 & 9.82 & 4.55 \\
L1448-mm Aperture 2 & 1.09 & 0.83--1.37 & 1.11 & 8.30 & 8.17--8.48 & 8.29 & 0.24 \\
L1448-mm Aperture 3 & 1.30 & 0.86--2.27 & 1.06 & 7.93 & 7.53--8.37 & 8.14 & 0.58 \\
L1448-mm Aperture 4 &  & $<$ 3.55 &  &  &  &  &  \\
Ser-emb 8 (N) Aperture 1 & 1.71 & 1.45--2.10 & 1.80 & 9.17 & 8.75--9.68 & 9.00 & 0.55 \\
Ser-emb 8 (N) Aperture 2 & 1.38 & 1.26--1.52 & 1.33 & 9.63 & 9.29--9.89 & 10.00 & 2.41 \\
\hline
\end{tabular}
\end{table*}

\section{Discussion}

\subsection{Quality of the fits to the F25 data}

Figure 3 shows that good fits can be obtained to the H$_2$ and HD
rotational diagrams obtained for BHR71 Aperture 2 and HH 211 Aperture 7, with reduced
$\chi^2$ of 1.16 and 1.34, respectively.  These values are not atypical, and equally
good fits are obtained for many other apertures, as shown in Table 3 where the
reduced $\chi^2$ values are shown in the rightmost column.  Two positions
show a reduced $\chi^2$ exceeding 5, both of which were observed toward NGC 1333 IRAS 4B.
This source is also an outlier with respect to $x({\rm HD})$, and the poor fits obtained
may indicate that the model (two isobaric temperature components with a common
HD abundance) is inadequate for this source.  The remaining 15 fits to
the apertures with HD detections have a median reduced $\chi^2$ of 1.34.  This may also
point to a shortcoming in the model or alternatively suggest
that the observational errors have been modestly underestimated (by $\sim 15 - 20\%$).

The corner plots shown in Figures 4 and 5 reveal degeneracies (covariances) between
particular pairs of model parameters.  These are indicated by elongated and
slanted contours in the two-dimensional panels.  
Two degeneracies are particularly pronounced and noteworthy.
First, the temperature and column density of the hot component are highly
degenerate with a negative covariance; this degeneracy occurs because the hot
component is primarily constrained by the most highly-excited H$_2$ states, for
which the model predictions are an increasing function of both temperature and column
density.  Second, for pressures $p/k_B < 3 \times 10^9\rm \,K\,cm^{-3}$, the pressure
and HD abundance are also degenerate with a negative covariance; this behavior arises
because, at densities sufficient to thermalize H$_2$ but not to thermalize
HD, the predicted excited-state HD column densities are approximately proportional to
the product $p\,x({\rm HD})$.

\Needspace{5\baselineskip}
\subsection{Derived physical parameters}

Figures 6 and 7 show a histogram and cumulative distribution function (CDF) for
the best-fit values obtained for each model parameter.   The sample comprises all
26 apertures presented by F25, except in the case of $\log(p/k_B)$ and $x({\rm HD})$, where
only the 17 apertures with HD detections are included.  The typical values obtained for
$N_w$, $N_h$, $T_w$, $T_h$, ${\rm OPR}_w$ and ${\rm OPR}_h$ are generally consistent
with previous results obtained from Spitzer and JWST
\citep{Neufeld2006a,Neufeld2007,Giannini2011,Francis2025,Neufeld2024}.
For the warm component, the best-fit
temperature, $T_w$, has a median value of 623~K, the best-fit column density,
$N_w$, has a median value of $1.5 \times 10^{20}\rm \, cm^{-2}$, and the
best-fit ortho-to-para ratio,
${\rm OPR}_w$, has a median value of 1.9.  The hot component,
accounting for 2.7$\%$ of the total
H$_2$ column on average, is considerably hotter with a median best-fit temperature, $T_h$,
of $\sim 2000$~K.  Its median OPR is 3.0, consistent with the equilibrium value expected
at temperatures above $\sim 300$~K.  The difference between the typical OPR$_w$
and OPR$_h$ has been widely noted in previous work
\citep{Neufeld2006a,Giannini2011} and attributed to the fact
that para-to-ortho conversion is slow, particularly at the temperature of the warm gas;
thus, below-equilibrium values for ${\rm OPR}_w$ are thought to be a relic of an earlier period,
prior to shock-heating, in which the OPR had equilibrated to a value characteristic
of cold molecular gas.
The best-fit values, median posterior values, and 68\% credible ranges for every parameter
and position are available as CSV files on Zenodo
\citep[doi:10.5281/zenodo.20663618]{NeufeldData2026}.

\begin{figure*}[p!]
\centering
\fullpagefig{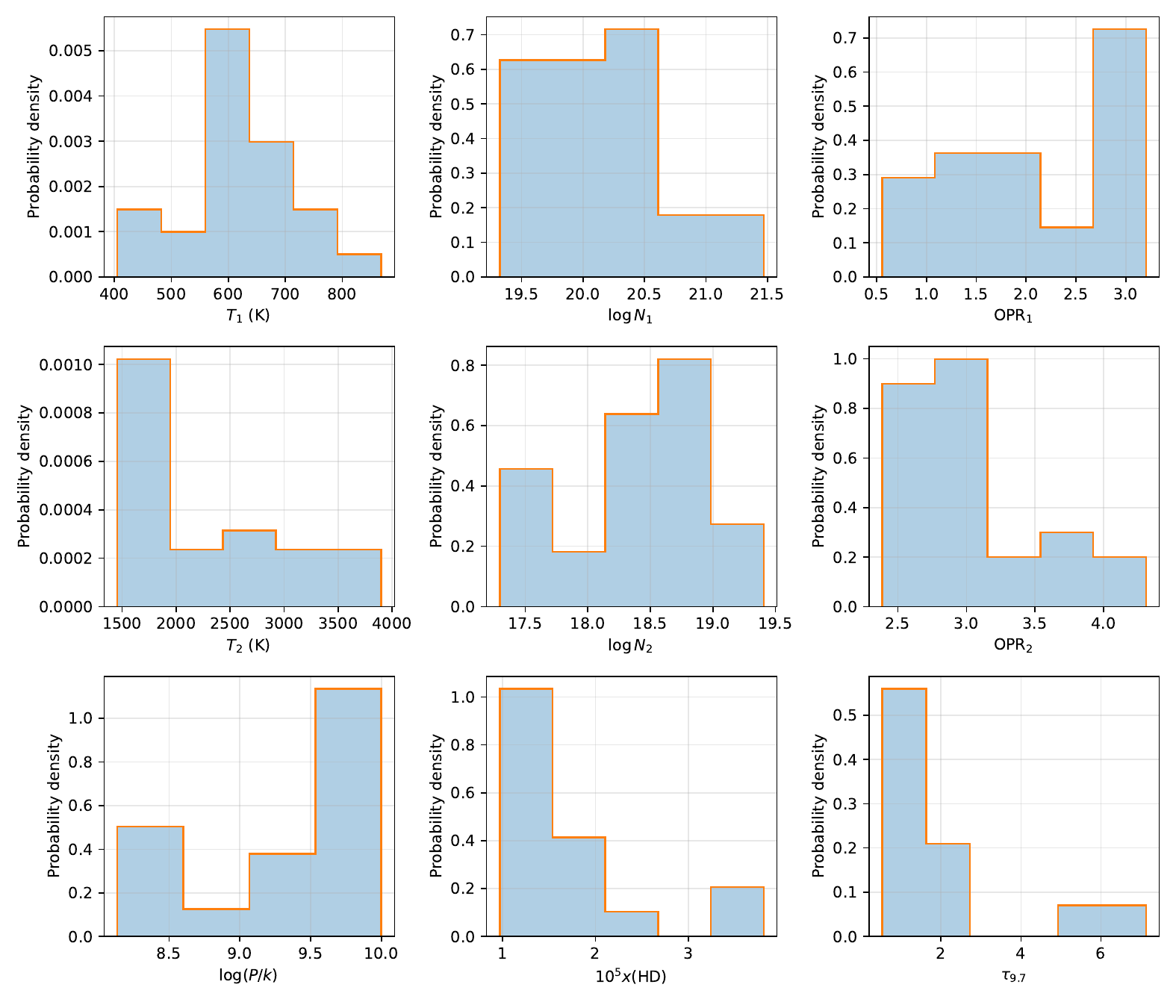}
\caption{Distributions of fitted physical parameters across the source sample.}
\label{fig:param-histograms}
\end{figure*}

\begin{figure*}[p!]
\centering
\fullpagefig{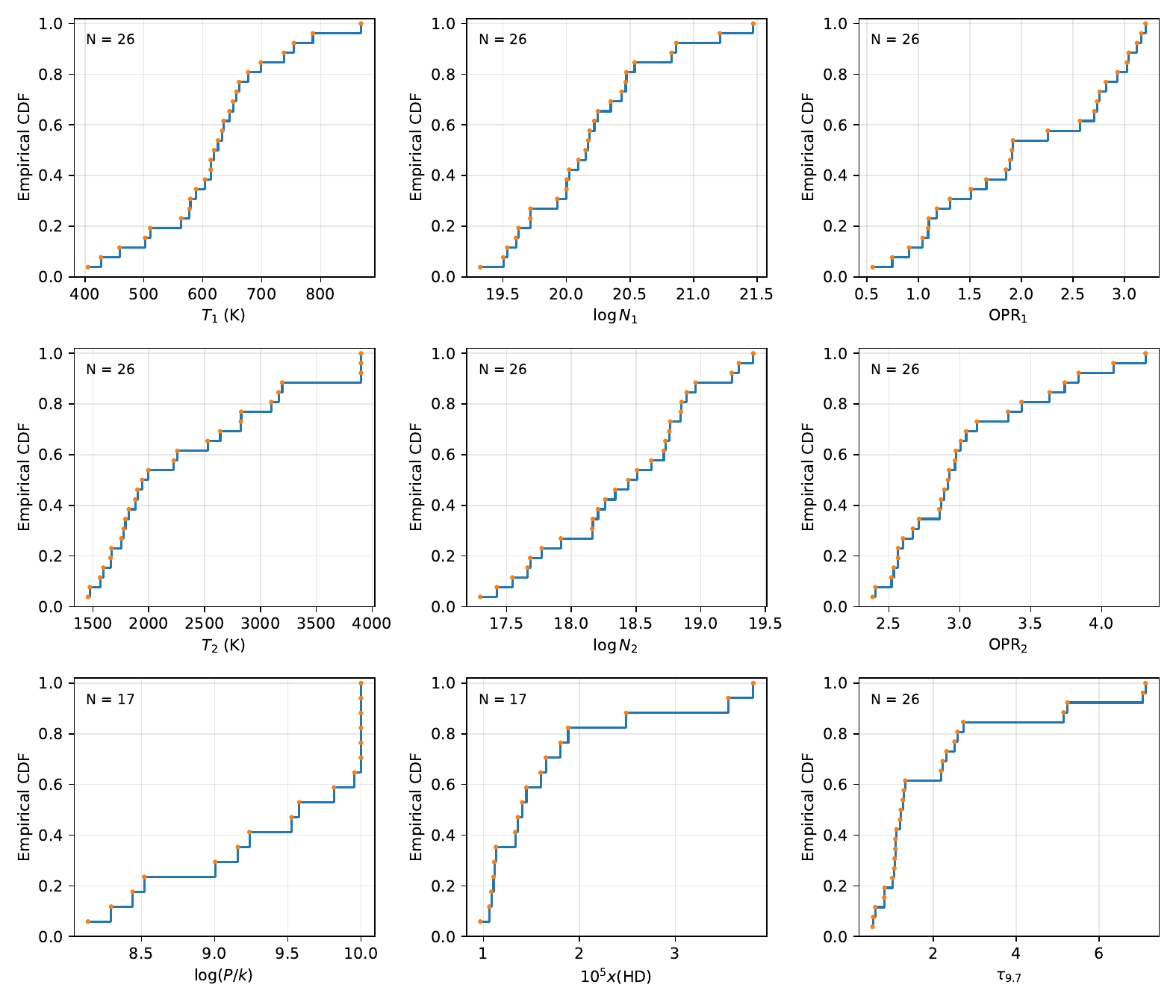}
\caption{Cumulative distribution functions for the fitted parameters.}
\label{fig:param-cdfs}
\end{figure*}

\begin{figure*}[p!]
\centering
\fullpagefig{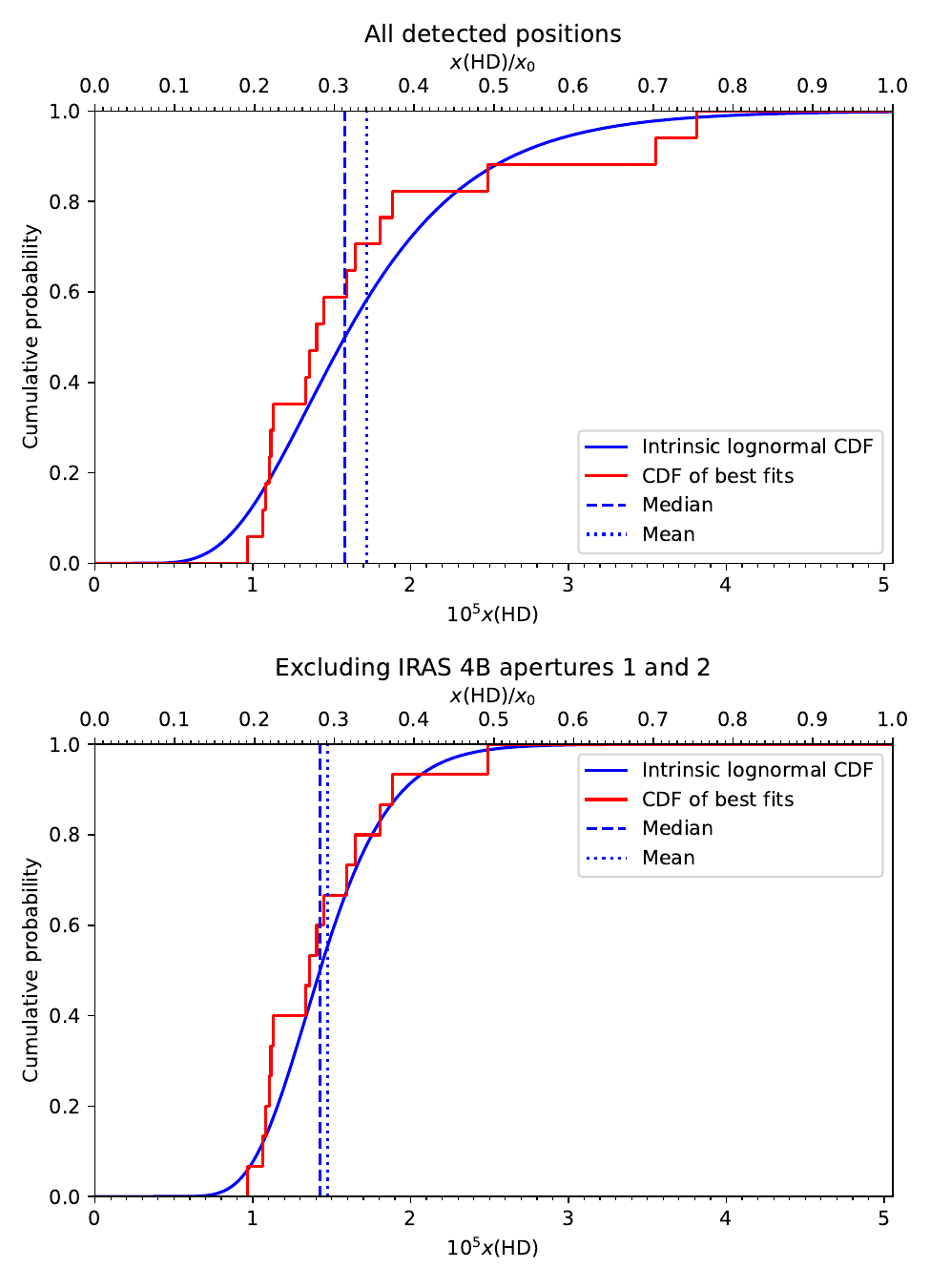}
\caption{Cumulative distribution of the HD/H$_2$ abundance ratio.  Red: CDF of
best-fit values.  Blue: estimate of the
intrinsic CDF, assumed log-normal, after accounting for uncertainties in the fitted values.}
\label{fig:intrinsic-xhd-cdf}
\end{figure*}

For the 17 apertures in which HD is also detected, the pressure, $p$,
and HD/H$_2$ abundance ratio, $x({\rm HD})$, are also constrained.
For 6 of the 17 apertures, the best-fit pressure is pinned to the upper limit
of the range considered ($p/k_B = 10^{10}\,\rm K\,cm^{-3}$).  In these cases, the
measured H$_2$ and HD level populations are consistent with sufficiently high pressures
that the HD line ratios are close to LTE and have lost most of their pressure sensitivity:
here the analysis yields only a lower limit on the pressure.  In the other 11 apertures, the
$p/k_B$ values range from  $8.4\times10^7$ to $5.1\times10^9\ {\rm K\,cm^{-3}}$ (median
posterior values).
It should be emphasized that these values apply to the emitting gas, which has
been pressurized by a shock, not the preshock material through which the shock is
propagating.  Although the pressures are high, they are not unreasonable for strong
shocks traveling through relatively dense material, particularly if the magnetic
field is relatively weak or is oriented along the shock-propagation direction.  Moreover, there
is likely a selection effect at play: the gas pressure can only be determined from
these data when HD is detected, but HD is only detectable if the pressure is large enough.

\subsection{Derived HD abundances}

Figure 8 shows the CDF for the HD/H$_2$ abundance ratio, $x({\rm HD})$, for
the 17 apertures with HD detections.  Here, the red ``staircase'' shows the
best-fit values, while the blue curve provides an estimate of the
intrinsic CDF.  The latter assumes a log-normal distribution for $x({\rm HD})$,
and accounts for the fact that the derived values for each aperture have
uncertainties.  In each panel, the lower horizontal axis shows $10^5 \times x({\rm HD})$.
The upper axis shows $x({\rm HD})/x_0$, where $x_0$ is the value that would be obtained
if all D were in HD, all H were in H$_2$, and the elemental abundance were equal to the
primordial value of ${\rm D/H}=(2.527\pm0.030)\times10^{-5}$ \citep{Cooke2018}.  Because H$_2$ contains two H
nuclei, whereas HD contains only one D nucleus, $x_0$ is twice the primordial elemental
abundance ratio.  In the upper panel, results are shown for all 17 apertures with
HD detections, while in the lower panel the positions toward IRAS 4B,
for which the fits are less reliable, are excluded.
After excluding the two IRAS 4B apertures, the median and mean values of $x({\rm HD})$
are $1.42\times10^{-5}$ and $1.47\times10^{-5}$, corresponding to 28\% and
29\% of $x_0$, respectively.  The dispersion of the best-fit values of
$\log_{10} x({\rm HD})$ is 0.17 dex for all 17 detections and decreases to
0.11 dex when the two IRAS 4B apertures are excluded, corresponding to
scatters of approximately 40\% and 30\%, respectively.
Figure 8, and the median and mean values
of $x({\rm HD})$ stated above, ignore non-detections of HD.  These non-detections do little
to modify the statistics, because 7 of the 9 upper limits are larger than the
median abundance (and also larger than any of the best-fit $x({\rm HD})$
values determined for apertures with HD detections outside IRAS 4B).

The present results agree qualitatively with the principal conclusion of F25 that the
inferred abundance of HD relative to H$_2$ is generally well below the value expected if
all primordial deuterium were present as HD.  Quantitatively, however, the derived
abundances differ substantially for several apertures because of the different
approaches adopted in the two studies: a detailed comparison of the two analyses is presented in
Appendix~\ref{app:previous-comparison}.  Moreover, whereas
F25 applied corrections to infer elemental D/H and discussed depletion of deuterium onto
dust and the destruction of HD in shocks, the quantity determined directly here
is $x({\rm HD})$; translating it into the gas-phase D/H ratio requires a detailed
analysis of shock chemistry that is beyond the scope of this study.

\section{Summary}

I have presented non-LTE excitation calculations for the rotational states of
HD and H$_2$ and used them to develop a method for fitting the two rotational
diagrams simultaneously.  The main results are as follows.

\begin{enumerate}

\item Because the dipole-allowed rotational transitions of HD have much larger
spontaneous radiative rates than the quadrupole transitions of H$_2$, HD
remains subthermally excited to substantially higher densities.  The H$_2$
rotational diagram therefore primarily constrains the temperature distribution
and warm-gas column, while the HD diagram provides complementary constraints
on density and pressure.

\item The joint non-LTE method was applied to JWST/MIRI measurements toward 26
positions in 10 protostellar outflows from the JOYS survey.  Two-temperature,
isobaric models generally reproduce the observed H$_2$ and HD
level populations well.  Excluding two positions toward IRAS 4B, the 15
apertures with HD detections have a median reduced $\chi^2$ of 1.34.

\item For 11 of the 17 apertures with HD detections, the inferred median
pressures span $p/k_B=8.4\times10^7$ -- $5.1\times10^9\ {\rm K\,cm^{-3}}$.
For the other six apertures, the HD rotational populations approach LTE and
provide only lower limits on the pressure.

\item After excluding the two comparatively poor fits toward IRAS 4B, the median
and mean inferred HD/H$_2$ abundance ratios are $1.42\times10^{-5}$ and
$1.47\times10^{-5}$, respectively.  These values are only 28\% and 29\%
of the ratio expected if all primordial deuterium were present as HD.

\end{enumerate}

The quantity directly constrained by this analysis is the HD/H$_2$ abundance
ratio, not elemental D/H.  Interpreting the low inferred HD abundances in terms
of astration, depletion onto dust, or preferential destruction of HD requires
additional modeling of the shock chemistry.

\begin{acknowledgments}
This work is based on observations made with the NASA/ESA/CSA James Webb Space Telescope.
I gratefully acknowledge support from JWST Theory grant JWST-AR-06829,
``Understanding the rich molecular spectra observed with NIRSpec and MIRI/MRS
toward protostellar outflows,'' and JWST grant JWST-GO-05804,
``HEFE: High Angular Resolution observations of Stellar Emergence in
Filamentary Environments.''  OpenAI Codex (GPT-5; \citealt{OpenAICodex})
was used to assist with software development, manuscript proofreading, and
LaTeX formatting.
\end{acknowledgments}

\facilities{JWST}

\software{corner, emcee, lmfit, Matplotlib, NumPy, OpenAI Codex (GPT-5), pypdf, SciPy}

\appendix

\section{Extension of the HD collisional rate coefficients}
\label{app:hd-rates}

The HD--H$_2$ calculations of \citet{Wan2019} provide downward state-to-state rate
coefficients for states up to $J=8$.  To model observed transitions with $J_U$ as
high as 10, and to avoid an artificial boundary immediately above the highest
levels constrained by MIRI, I extended the rate matrix through $J=14$.  The extension was
performed separately for ortho- and para-H$_2$ collision partners and independently at each
of the temperatures 100, 200, 300, 400, 600, 1000, 2000, 3000, and 4000~K.

\begin{figure*}[p!]
\centering
\fullpagefig{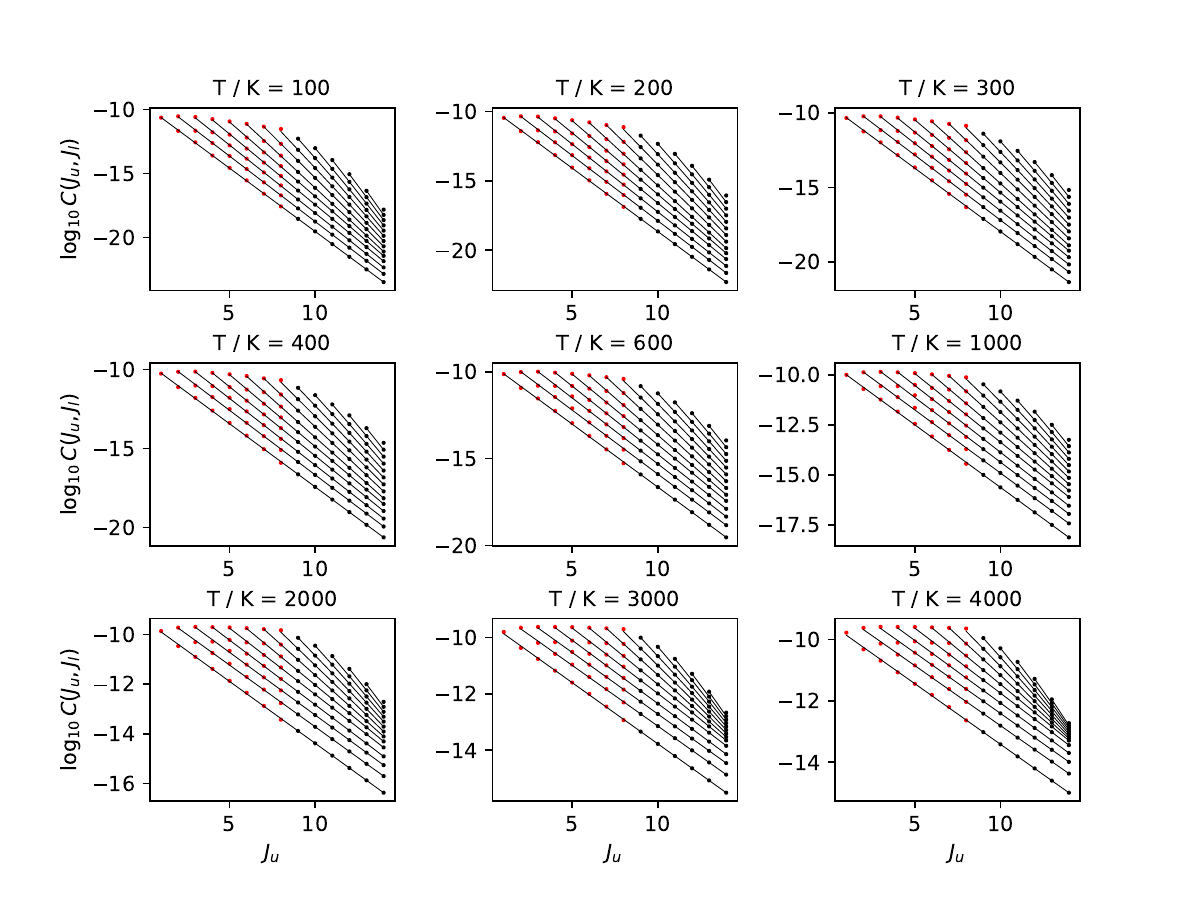}
\caption{HD collisional rate coefficients for collisions with ortho-H$_2$.}
\label{fig:app-rates-j-extra-oh2}
\end{figure*}

At fixed temperature and lower state $J_L$, the available downward rates were fitted as
\[
\log_{10} C(T,J_U,J_L)=b(T,J_L)+a(T,J_L)J_U .
\]
This linear dependence of $\log_{10} C(T,J_U,J_L)$ on $J_U$
provides a good fit to the \citet{Wan2019} results, which are
denoted by red points in Figure \ref{fig:app-rates-j-extra-oh2}.  Each black line passing
through a set of red points shows the best linear fit to the data for a given $J_L$,
which increases monotonically from the bottom ($J_L=0$) to the top.
The fitted relation was then evaluated
for the missing upper states to obtain the black points.
For lower states beyond the range that permits a direct
fit, the slope and intercept were continued using
their local trend between the last two fitted lower states\footnote{As an alternate extrapolation 
method, I adopted $\log_{10} C(T,J_U,J_L)=b^\prime(T,J_L)+a(T,J_L)(J_U-J_L-1)$ with
$b^\prime(T,J_L)$ extrapolated linearly for $J_L \ge 8$.  This expression yields a
smoother behavior for $\log_{10} C(T,J_U,J_L)=b^\prime(T,J_L)$ but leads to only
negligible changes in the values of $p$ and $x(HD)$ inferred from astronomical observations 
in Section 3.3}.
Upward coefficients were obtained from detailed balance.
The extension is entirely empirical: it cannot reproduce
unanticipated high-$J$ resonances or changes in propensity rules.  Its purpose is to extend
the statistical-equilibrium equations above the observed levels, not to provide new
quantum-scattering data.  The results are provided in a LAMDA-style file on Zenodo
\citep[doi:10.5281/zenodo.20663618]{NeufeldData2026}.  

\section{Details of the fitting procedure}
\label{app:fitting-details}

For each aperture, the fitted data are the column densities, $N_J$, inferred
in the optically thin limit from the H$_2$
and HD line fluxes measured by F25.  A 6\% calibration uncertainty
is added in quadrature to the reported statistical uncertainty of every line.

The model column densities, $N_J$, for each species are the sum of warm and hot components.  For H$_2$,
each component has a fitted total column, kinetic temperature, and OPR.  The two components
share a pressure, so their collider densities are $n({\rm H_2})=(p/k_B)/T_{\rm kin}$.
The HD columns are the corresponding H$_2$ columns multiplied by the single fitted
$x({\rm HD})$.  Population fractions are interpolated within the non-LTE grids described in
Section 2.  Thus the nine fitted parameters are
\[
\{\log N_w,T_w,{\rm OPR}_w,\log N_h,T_h,{\rm OPR}_h,
\log(p/k_B),x({\rm HD}),\tau_{9.7}\}.
\]

\begin{figure*}
\centering
\fullpagefig{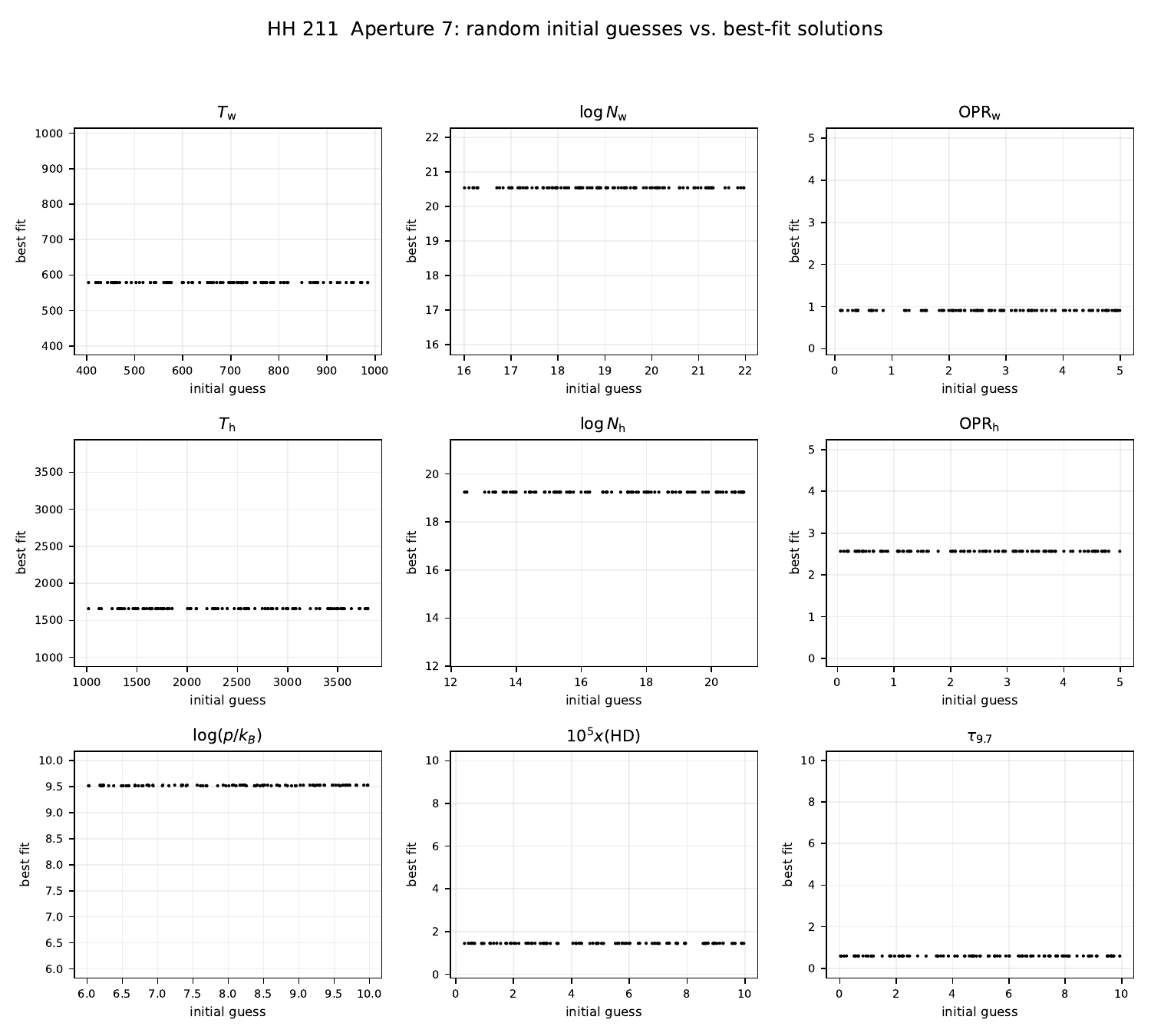}
\caption{Least-squares results for HH 211 Aperture 7. Each panel compares the random
initial guess for one fitted parameter with the best-fit value reached by that
least-squares run.}
\label{fig:app-hh211-ap7-multistart}
\end{figure*}

The first fitting method uses a bounded trust-region reflective least-squares
algorithm, as implemented by {\tt lmfit},
to determine the model parameters that minimize $\chi^2$.
This method sets bounds on all fitted parameters
and requires an initial guess of the best-fit values.
To investigate any possible dependence on that initial guess---such
as might result from the existence of multiple local minima in $\chi^2$---I selected
100 random starting points within the bounded
nine-dimensional parameter space.  For every aperture, all tested starting points
converged to the same solution, suggesting that the global minimum was found.
Example results, obtained for HH 211 Aperture 7, are shown in Figure \ref{fig:app-hh211-ap7-multistart},
where the solution is plotted as a function of the initial guess.

The second fitting method uses MCMC to sample the nine-dimensional parameter space.
The posterior is sampled with the affine-invariant ensemble method of
\citet{GoodmanWeare2010}, using the implementation of \citet{ForemanMackey2013}
provided by {\tt emcee}.  For each aperture, 20 walkers were initialized in a Gaussian
cloud around the least-squares solution, with a standard deviation of $10^{-4}$ times
the value of each parameter.  Each walker was initially advanced for 80,000 steps, after
which convergence was checked every 10,000 steps up to a maximum of 200,000 steps.
The burn-in length was the larger of 10,000 steps and five times the largest estimated
integrated autocorrelation time, $\tau$.  Sampling was stopped when the post-burn-in chain
exceeded $50\tau$ for every parameter, the effective sample size exceeded 1000 for every
parameter, and successive estimates of $\tau$ agreed to within 5\%.  These criteria were
met for 23 of the 26 apertures; HH 211 Apertures 4--6 reached the 200,000-step limit.
For posterior summaries and subsequent analysis, 70,000 evenly spaced post-burn-in
samples were retained per aperture.  The priors are uniform within the hard bounds
listed in Section 3.2 and zero outside them.
For apertures without a secure HD detection, the HD abundance is reported as an upper limit
from the posterior rather than as a measurement.

\begin{figure*}[p!]
\centering
\fullpagefig{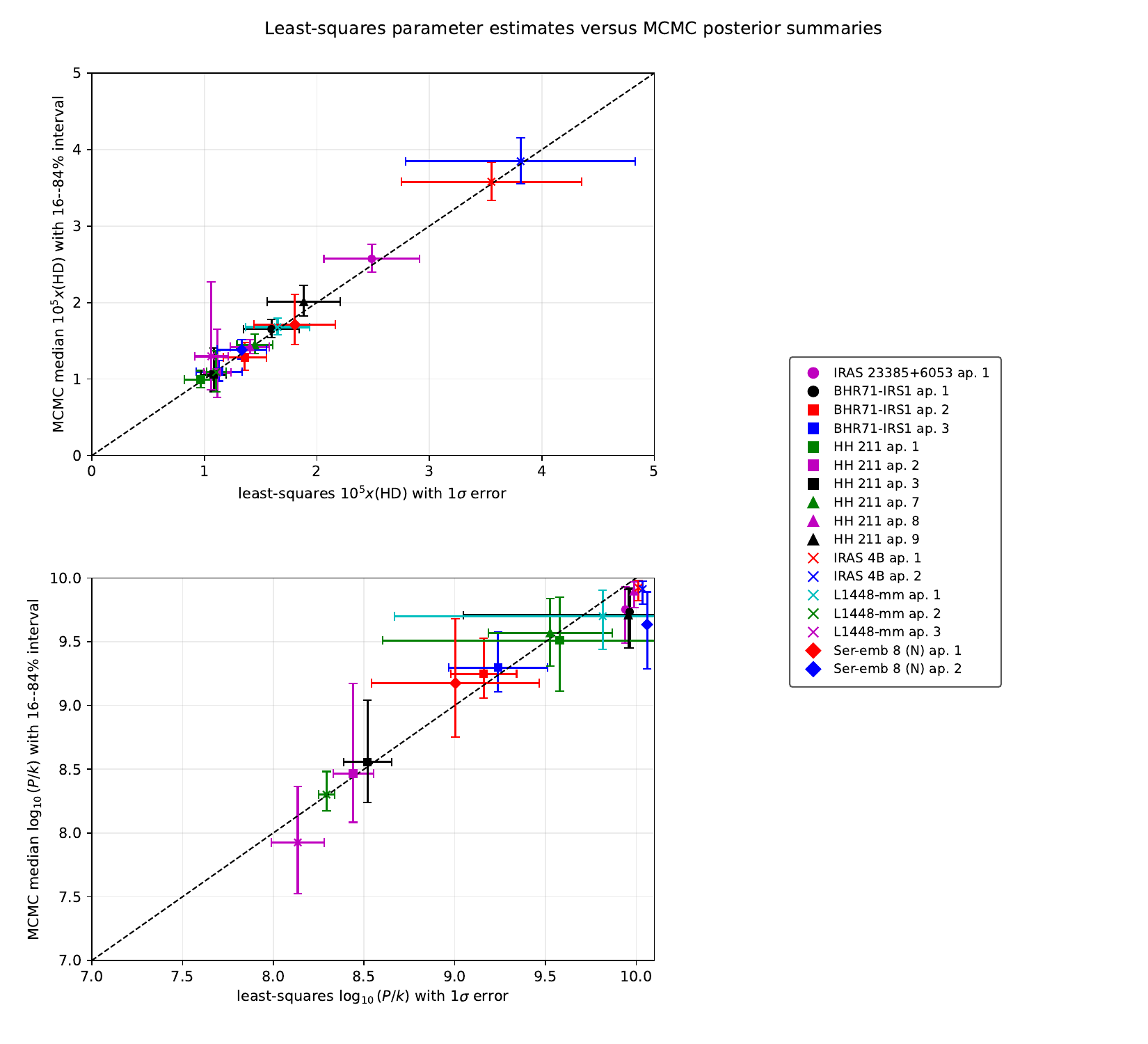}
\caption{Comparison of least-squares parameter estimates with the corresponding MCMC
estimates for the HD/H$_2$ abundance ratio and gas pressure. Points show the
least-squares best values with formal $1\sigma$ errors against the MCMC medians with
central 68\% credible intervals.}
\label{fig:app-best-vs-mcmc}
\end{figure*}

Figure \ref{fig:app-best-vs-mcmc} compares the least-squares estimates with the MCMC
summaries.  Agreement is generally good for $x({\rm HD})$.  The pressure is more susceptible to
asymmetry because the line ratios saturate near LTE.  This is why the
MCMC results are adopted for interval estimates even though the least-squares solutions are
useful diagnostics.

\begin{figure*}[p!]
\centering
\fullpagefig{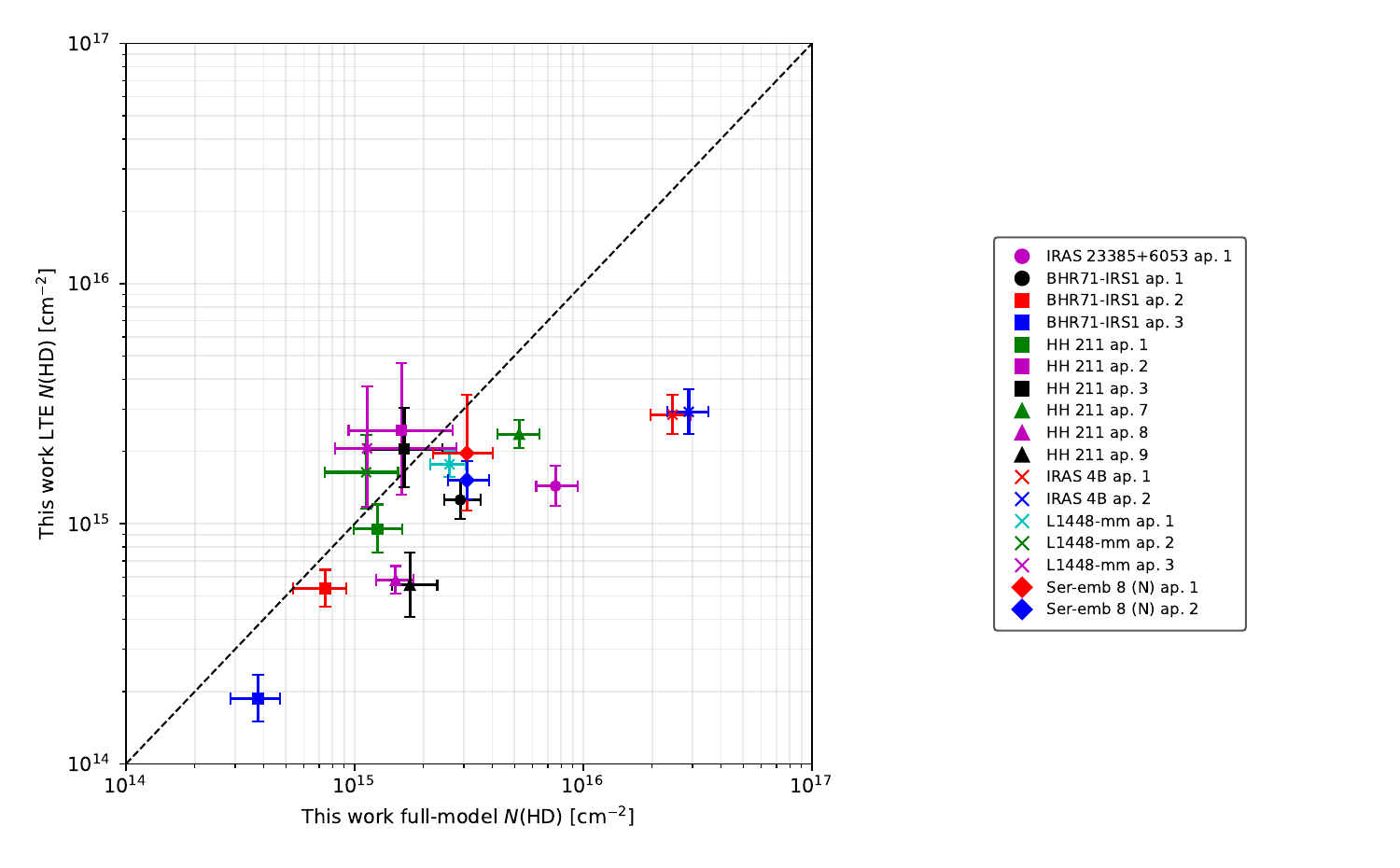}
\caption{Comparison of LTE HD rotational-diagram column densities with the full non-LTE
HD column estimates from the present model.}
\label{fig:app-hd-lte-vs-full}
\end{figure*}

The total HD column densities derived from simultaneous non-LTE fits to the
HD and H$_2$ rotational diagrams can differ substantially from those obtained
from an LTE fit to the HD diagram alone.  Figure \ref{fig:app-hd-lte-vs-full} shows that
comparison for the 17 positions with detected HD.  The differences arise
almost entirely from how the rotational diagrams extrapolate to the
lowest-$J$ states ($J=0$--4) that cannot be observed with MIRI.

\begin{figure*}[p!]
\centering
\fullpagefig{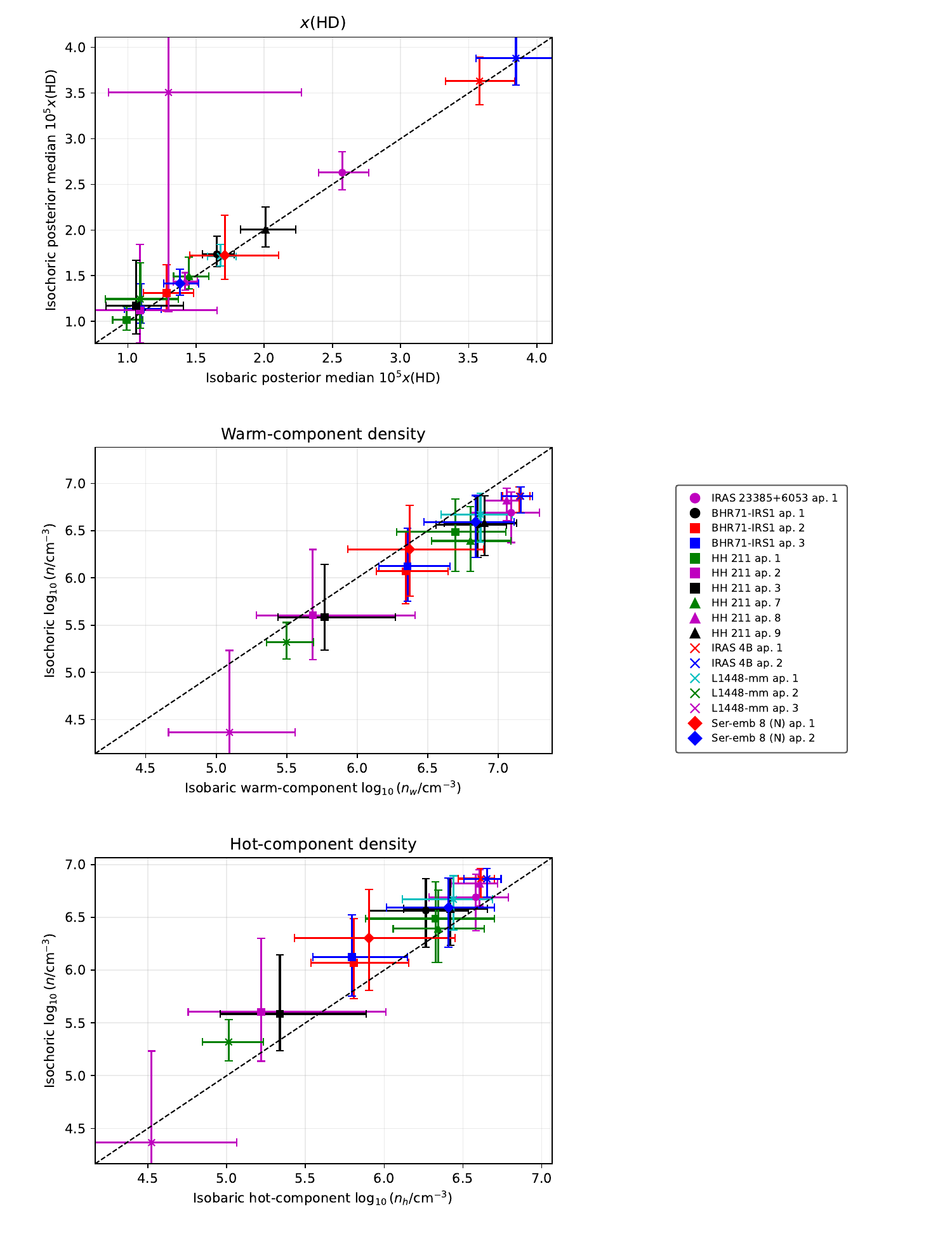}
\caption{Comparison of the isobaric and isochoric posterior results for positions
with detected HD.  The upper panel compares the inferred HD abundances, while
the middle and lower panels compare the isochoric density with the warm- and
hot-component densities, respectively, in the isobaric model.  Points show
posterior medians with central 68\% credible intervals.}
\label{fig:app-isoc-vs-isobaric}
\end{figure*}

As discussed in Section 3.2, Figure \ref{fig:app-isoc-vs-isobaric} compares the 
posterior results obtained with the standard model to those obtained with an alternative
model in which the warm and hot components are assumed isochoric instead of isobaric.
The inferred HD abundances are similar, while the single density in the
isochoric model generally lies between the warm- and hot-component densities
implied by the common-pressure isobaric model.  The isochoric model will not be considered
further.

Although the OPR of the emitting H$_2$ is a free parameter, the ortho- and para-H$_2$
collisional rate coefficients were combined assuming a collider OPR of 3 for computational
simplicity.  To estimate the sensitivity to this assumption, I compared effective
de-excitation rate coefficients computed for collider OPRs of 1 and 3, retaining transitions
for which both coefficients exceed $10^{-15}$~cm$^3$~s$^{-1}$.  At 500~K, the OPR = 1
effective rates are smaller by $3\pm2\%$ with HD as the target and $8\pm4\%$ for H$_2$,
where the quoted values are the mean and standard deviation.  Similar differences are
obtained at 1000~K.  This comparison is conservative because the fitted warm-component OPRs
do not fall below 1, while the hot-component OPRs are usually consistent with 3.  Thus,
adopting a collider OPR of 3 likely introduces only a small uncertainty in the excitation
calculation.

\begin{figure*}[p!]
\centering
\fullpagefig{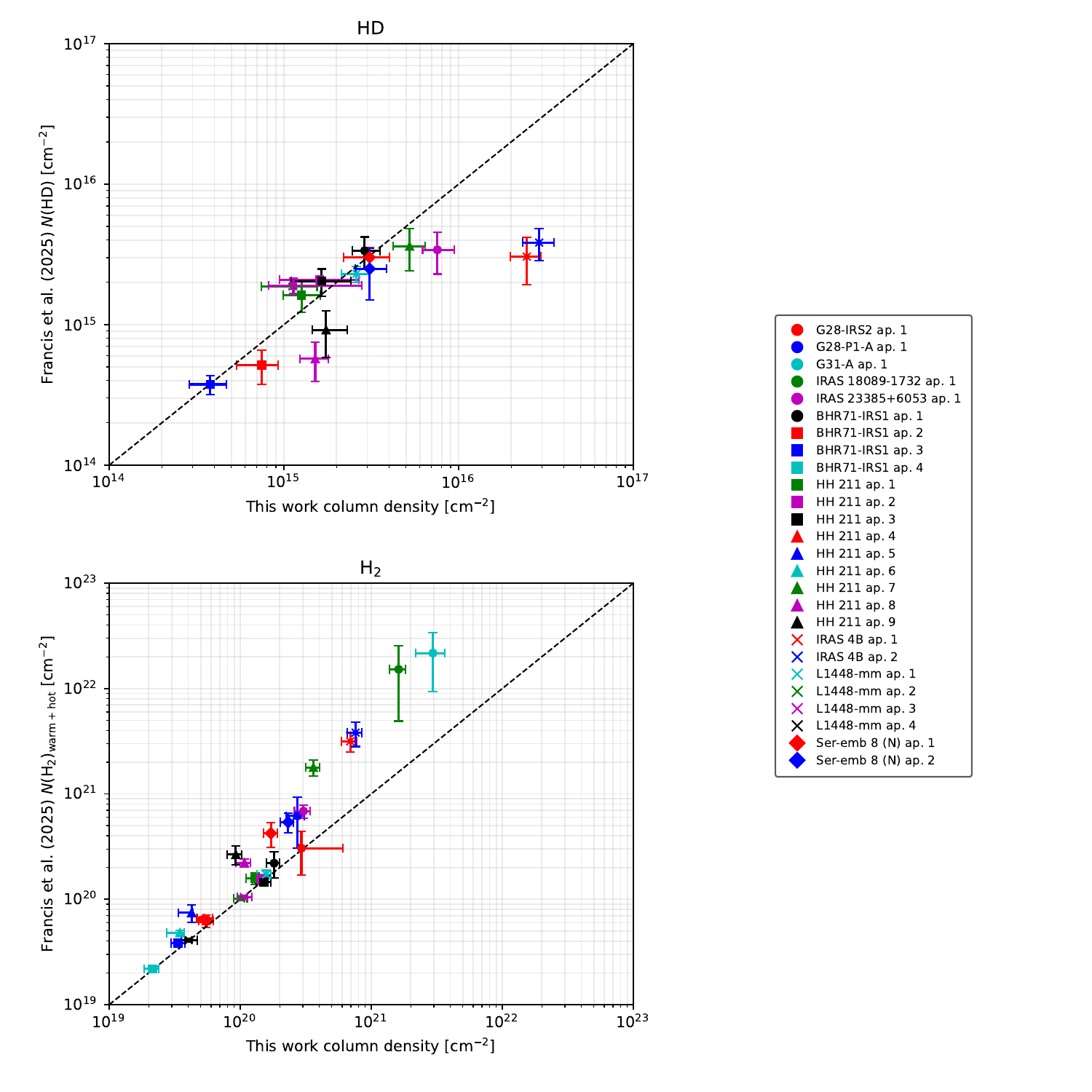}
\caption{Comparison of HD and H$_2$ column densities from the present non-LTE analysis
with those from F25. The upper panel shows the HD comparison and
the lower panel shows the H$_2$ comparison.}
\label{fig:app-francis-columns}
\end{figure*}

\section{Detailed comparison with previous results}
\label{app:previous-comparison}

Figure \ref{fig:app-francis-columns} compares the results obtained for
the total H$_2$ and HD column densities in the present work with those
presented by F25.  Although the results for HD agree well for many positions
(upper panel), there is a tendency for the present analysis to yield larger HD
column densities.  These differences result from my inclusion of non-LTE excitation
and the assumption that the temperature structure of the HD-emitting gas
mirrors that of the H$_2$-emitting gas.

Similarly, the H$_2$ column densities agree well for many positions (lower panel),
but for H$_2$ there are multiple cases where the column densities derived here
are a factor of several smaller than those obtained by F25.  These are the more highly
extinguished sources, for which the extinction correction computed by F25 is
systematically larger than that obtained in the present study.  That systematic effect,
in turn, is likely traceable to different assumptions about the OPR.  F25 adopted a common OPR
for both the warm and hot components, whereas in the present study I adopted separate
values for ${\rm OPR}_w$ and ${\rm OPR}_h$.  Because ${\rm OPR}_w$ is typically smaller
than ${\rm OPR}_h$, as discussed in Section 4.2, the values for ${\rm OPR}_w$ obtained
here are smaller than the common OPR obtained by F25.  This leads to a smaller model prediction
for the unreddened H$_2$ S(3) flux, and because the S(3) transition is the most strongly
extinguished among the fitted H$_2$ lines, a smaller estimate of the extinction.

\end{document}